\documentclass[english,aps,prl,twocolumn,amsmath,amssymb,letterpaper,showpacs,superscriptaddress]{revtex4-1}
\usepackage{amsmath}
\usepackage{amssymb}
\usepackage{nicefrac}
\usepackage{graphicx}
\usepackage{pdfpages}	
\usepackage{pgffor}		
\usepackage[unicode=true,pdfusetitle,
 bookmarks=true,bookmarksnumbered=false,bookmarksopen=false,
 breaklinks=false,pdfborder={0 0 1},backref=false,colorlinks=false]
 {hyperref}

\makeatletter
\date{May 2, 2016}
\makeatother

\begin{document}

\title{Magnetic moments in a helical edge can make weak correlations seem strong}

\author{Jukka I. V\"ayrynen}

\affiliation{Department of Physics, Yale University, New Haven, CT 06520, USA}

\author{Florian Geissler}

\affiliation{Institute for Theoretical Physics and Astrophysics, University of W\"urzburg, 97074 W\"urzburg, Germany}

\author{Leonid I. Glazman}

\affiliation{Department of Physics, Yale University, New Haven, CT 06520, USA}

\begin{abstract}
We study the effect of localized magnetic moments on the conductance of a helical edge. Interaction with a local moment is an effective backscattering mechanism for the edge electrons. We evaluate the resulting differential conductance as a function of temperature $T$ and applied bias $V$ for any value of $V/T$. Backscattering off magnetic moments, combined with the weak repulsion between the edge electrons results in a power-law temperature and voltage dependence of the conductance; the corresponding small positive exponent is indicative of insulating behavior. Local moments may naturally appear due to charge disorder in a narrow-gap semiconductor. Our results provide an alternative interpretation of the recent experiment by Li et al.~\cite{Li15} where a power-law suppression of the conductance was attributed to strong electron repulsion within the edge, with the value of Luttinger liquid parameter $K$ fine-tuned close to $1/4$.
\end{abstract}

\maketitle

{\it Introduction -} In search for topological insulators, the III-V semiconductor structures with band inversion appeared as a viable option~\cite{liu08}. The band inversion does occur in the type-2 heterostructure, InAs/GaSb. If the layers forming the well are narrow enough, the hybridization of states across the interface results in a formation of a gap; in the ``topological'' phase, the gap is accompanied by edge states free from elastic backscattering. These putative states became a target of an extensive set of measurements~\cite{Knez14,Spanton04,Du2015,Li15,Nichele2015}. First, a surprisingly robust conductance quantization was found~\cite{Du2015}.  
A later experiment~\cite{Li15} explained the temperature-independent quantized conductance $G$ as an inadvertent deviation from the linear-response regime. 
 The observed~\cite{Li15} power-law temperature and bias voltage dependence of the differential conductance was suggestive of insulating behavior.
Assuming topologically protected edge states, it can be
interpreted as a manifestation of strong-interaction physics: at
low energies, even a single impurity can ``cut'' the edge, 
suppressing charge transport~\cite{KFPRL92} if the Luttinger parameter is very small, $K<1/4$~\cite{wu_helical_2006,Xu06} ($K=1$
corresponds to non-interacting electrons). Measurements~\cite{Li15} yield $K\approx 0.22$ (with a 5\% error), which is very close to the critical value of $1/4$; an increase of $K$ by mere 12\% would change the sign of $dG/dT$.
Fine-tuning $K$ to such a stable value seems improbable, given the
dependence of the edge state velocity on the 
gate voltages, varied in the experiment. 
The reliance on fine-tuning in the current explanation of experiments  provides an impetus to search for alternatives less sensitive to a specific value of $K$.


We find that scattering off localized magnetic moments may lead to temperature and bias dependences of the differential conductance similar to those observed~\cite{Li15}  at moderately
weak interaction, $K\approx0.8$, without fine-tuning of $K$. The origin
of localized moments in InAs/GaSb quantum wells is not known, but the
narrow 40-60K gap in these systems may allow for the presence of
charge puddles~\cite{vayrynen_helical_2013} which can act as magnetic
impurities~\cite{vayrynen_resistance_2014}. 
In the present work we focus on the non-linear current-voltage characteristics and on the effects of electron-electron interactions within the helical edge which were not considered in Ref.~\cite{vayrynen_resistance_2014}. 


{\it The setup and qualitative description of the main results -} 
We start by considering a single spin-1/2 magnetic moment $\mathbf{S}$
coupled to a helical edge.
The isolated edge is described~\cite{wu_helical_2006} by a Luttinger liquid Hamiltonian $H_{0}$;
the local moment is coupled to the edge electrons by, generally, anisotropic exchange interaction.
Separating out its isotropic part, 
the full time-reversal symmetric Hamiltonian of the coupled
edge-impurity system can be written as 
\begin{equation}
H=H_{\text{iso}}+\sum_{ij}\delta J_{ij}S_{i}s_{j}(x_{0})\label{eq:HamiltonianFull}
\end{equation}
with $H_{\text{iso}}$ being the Hamiltonian with isotropic exchange:
\begin{equation}
H_{\text{iso}}=H_{0}+J_{0}\mathbf{S}\cdot\mathbf{s}(x_{0})\label{eq:HamiltonianIso}
\end{equation}
Here, $\mathbf{S}$ is the spin-1/2 impurity spin operator, and $\mathbf{s}(x_{0})=\frac{1}{2}\sum_{\alpha\beta}\psi_{\alpha}^{\dagger}(x_{0})\boldsymbol{\sigma}_{\alpha\beta}\psi_{\beta}(x_{0})$
is the edge electron spin density at the position $x_{0}$ of the contact interaction with the local moment.
(From hereon we will omit the position arguments.) We shall assume
$\delta J_{ij}\ll J_{0}$ so that the exchange is almost isotropic~\cite{vayrynen_resistance_2014}.
Thus we can treat the second term in Eq.~\eqref{eq:HamiltonianFull}
as a perturbation. 

The first term in $H_{\text{iso}}$, Eq.~\eqref{eq:HamiltonianIso} is the bosonized Luttinger-liquid Hamiltonian describing the interacting
helical edge electrons, $H_{0}=(2\pi)^{-1}v\int dx[\Pi^{2}+(\partial_{x}\varphi)^{2}]$; we assume the dimensionless exchange coupling parameter to be small, $\rho J_{0}\ll1$ (here $\rho$ is the electron
density of states per spin per unit edge length).
The bosonic fields commute as $[\varphi(x),\Pi(y)]=i\pi\delta(x-y)$.
We have rescaled the fields by appropriate factors of $\sqrt{K}$;
the bosonization identity is $\psi_{\beta}(x)=(2\pi a)^{-1/2}e^{-i(\beta\sqrt{K}\varphi-\frac{1}{\sqrt{K}}\int_{-\infty}^{x}dx'\Pi)}$
with $\beta=+/-$ for right/left movers (or spin up/down; we take
$\mathbf{z}$-axis to be the spin quantization axis of helical electrons
at Fermi energy); $a$ is the short-distance cutoff. 
In bosonic representation, the spin density takes form $s_{x}\pm is_{y}=\pm i(2\pi a)^{-1}e^{\pm2i\sqrt{K}\varphi}$, $s_{z}=\frac{1}{2\pi\sqrt{K}}\Pi$. Using it, we re-write the exchange interaction Hamiltonian as
\begin{equation}
\!J_{0}\mathbf{S}\cdot\mathbf{s} \!\to\! J_{\perp}\frac{-i}{4\pi a}(S_{+}e^{-2i\sqrt{K}\varphi}-S_{-}e^{2i\sqrt{K}\varphi})+J_{z}\frac{1}{2\pi\sqrt{K}}S_{z}\Pi\,.
\label{eq:uniaxialanisotropy}
\end{equation}
Even though the bare Hamiltonian~(\ref{eq:HamiltonianIso}) is isotropic, $J_{\perp} = J_z = J_0$, the exchange becomes anisotropic under renormalization group (RG) flow, as the scaling dimensions of the corresponding spin densities in Eq.~(\ref{eq:uniaxialanisotropy}), $\Delta_{\perp}=K$ and $\Delta_{z}=1$, differ from each other~\cite{senechal}, see also Eqs.(\ref{eq:RGJperp})--(\ref{eq:RGJz}) below.
The isotropy breaking is not an artefact: anisotropy 
is already present in the bare Hamiltonian even at $K=1$ due to the spin-orbit interaction; the Hamiltonian has no SU(2) symmetry but
only a smaller U(1) symmetry (spin rotations about $z$-axis). 

The weak-coupling ($\rho J\ll1$ and $1-K\ll1$) RG equations for
$J_{\perp}$ and $J_{z}$ are~\cite{lee_kondo_1992,furusaki_kondo_1994,Maciejko2009}
(here $E$ is the running cutoff) 
\begin{flalign}
\frac{dJ_{\perp}}{d\ln E} & =-(1-K)J_{\perp}-\rho J_{z}J_{\perp}\label{eq:RGJperp}\\
\frac{dJ_{z}}{d\ln E} & =-\rho J_{\perp}^{2}\label{eq:RGJz}
\end{flalign}
The right-hand-side of the first equation starts at tree level with
a coefficient~\cite{cardy1996scaling,senechal} $1-\Delta_{\perp}=1-K$;
the second equation does not have such a term since $\Delta_{z}=1$.
The terms second-order in $J$ are due to the Kondo effect and can
be derived from poor man scaling~\cite{anderson_poor_1970}, or from
an operator product expansion~\cite{cardy1996scaling,senechal}. 

Starting from isotropic initial condition, $J_{0}>0$, Eq.~\eqref{eq:RGJperp}
shows that there are two regimes of parameters: $\rho J_{0}\ll1-K$
and $\rho J_{0}\gg1-K$. 
In the latter case $1-K$ can be dropped
from Eq.~\eqref{eq:RGJperp}, and the physics is similar to that of the case $K=1$~\cite{vayrynen_resistance_2014}. 

In this paper we focus on the opposite limit,
$\rho J_{0}\ll1-K$. (Note, such initial condition can be satisfied
even if the electron-electron interaction is weak, $1-K\ll1$.) In
this case the RG flow governed by Eqs.~\eqref{eq:RGJperp}--\eqref{eq:RGJz}
can be divided into two regimes separated by energy scale
$T^{*}$ (we use units $k_{B}=\hbar=1$) defined by the crossover condition~\cite{Giamarchi} $\rho J_{z}(T^{*})=1-K$,
\begin{equation}
T^{*}=D(\frac{1}{\sqrt{2}}\frac{\rho J_{0}}{1-K})^{1/(1-K)}\,.
\label{eq:TstarDef}
\end{equation}
Here $D\sim E_{g}$ is the bare cutoff which we take to be the bulk band gap~\footnote{In general, the cutoff depends on the microscopic structure of the
impurity. If the exchange term in Eq.~\eqref{eq:HamiltonianFull}
originates from a charge puddle, the cutoff is $D\sim\min(E_{g},\,E_{C},\,\delta)$, where $E_C$ and $\delta$ are, respectively, the energies of charged and chargeless excitations in a puddle~\cite{vayrynen_resistance_2014}. }. 
At energies $E\gg T^{*}$ one can ignore $\rho J_{z}(E)$ in~\eqref{eq:RGJperp},
whereas at $E\ll  T^{*}$ one can ignore $1-K$. Next, we discuss
electron backscattering in the high energy limit, $E\gg T^{*}$
where interaction ($K\neq 1$) is important.

{\it The backscattering current at energies above $T^*$ -}%
The isotropic exchange Hamiltonian~\eqref{eq:HamiltonianIso} alone
does not backscatter edge electrons in steady state (DC bias) since
each backscattering event is accompanied by an action of the nilpotent
operator $S_{-}$ on the impurity spin polarized along $\mathbf{z}$-axis~\cite{tanaka_conductance_2011}.
The presence of anisotropy in the exchange, Eq.~\eqref{eq:HamiltonianFull},
gives rise to backscattering. This perturbation in Eq.~\eqref{eq:HamiltonianFull}
can be treated using Fermi Golden Rule, assuming equilibrium impurity
polarization $\left\langle \mathbf{S}\right\rangle =\mathbf{z}\frac{1}{2}\tanh\frac{eV}{2T}$~\footnote{The bias voltage creates an imbalance between left and right movers and a non-zero \unexpanded{$ \langle s_{z} \rangle$}  in Eq.~(\ref{eq:HamiltonianIso}). This results in non-zero \unexpanded{$ \langle S_{z} \rangle$} through the exchange interaction~(\ref{eq:HamiltonianIso})}.
Integration over electron phase space volume leads to a backscattering
current $\left\langle \delta I\right\rangle \sim e^{2}V(\rho\delta J)^{2}$.
We can find the full temperature and bias voltage dependence by solving
for the renormalized coupling $\delta J$. Since the pertinent
constant $\delta J$ couples to the spin-flip operators $e^{\pm2i\sqrt{K}\varphi}$,
it acquires a power-law energy dependence $\delta J(E)=(D/E)^{1-K}\delta J(D)$
for $E>T^{*}$. Taking $E\sim\max(T,\,eV)$, the $T$ and $V$ -dependent
backscattering current becomes (valid at $\max(T,\,eV)\gg T^{*}$)
\begin{equation}
\left\langle \delta I\right\rangle =\frac{e^{2}}{h}cV\,T^{-2(1-K)}[\max(1,\,eV/T)]^{-2(1-K)}\,,
\label{eq:QualitCurrentFinal}
\end{equation}
where constant $c$ depends on the bare exchange tensor.
Equation~\eqref{eq:QualitCurrentFinal} is a simplified version of our main result.
Its detailed version, see Eq.~\eqref{eq:current3}, reveals, in addition to $eV/T\sim 1$, yet another crossover in the current-voltage characteristic occurring at
$\frac{eV}{T}\sim\rho J\ll1$; it is associated
with the details of impurity spin torque and relaxation, ignored in Eq.~(\ref{eq:QualitCurrentFinal}).

{\it Long edge conductance at energies above $T^*$ -} Let us now
consider a long sample which may host many impurities near the edge.
A single impurity contributes an amount $\delta R\approx\delta G/G_{0}^{2}$
to the edge resistance (here $G_{0}=e^{2}/h$ and $\delta G=d\left\langle \delta I\right\rangle /dV$). 
In a long sample with $N$ impurities we can simply add resistances if the impurities are dilute enough~\footnote{See Supplemental Material for details.  The supplement includes references to~\onlinecite{BreuerPetruccione,Gradshteyn,GiamarchiSchulz,Gornyi05,Dyakonov,maciejko_kondo_2012,eriksson_spin-orbit_2013}}. 
The impurities dominate the edge resistance if $N\delta G\gg G_{0}$,
where the same typical value $\delta G$ for each impurity is used. In this
case one finds $G\approx G_{0}^{2}/N\delta G$ for the conductance
of a single edge. Here $\delta G$ is evaluated with the help of Eq.~\eqref{eq:current3}
or its simplified version, Eq.~\eqref{eq:QualitCurrentFinal}, both
valid at $\max(T,\,eV)>T^{*}$.

Using 
Eq.~\eqref{eq:QualitCurrentFinal} one finds
a power-law dependence $G(V,T)\approx(G_{0}/cN)[\max(T,\,eV)]^{2(1-K)}$.
 In Ref.~\cite{Li15} the authors found a fit $G\propto V^{0.37}$
in the regime $eV>T$ for
a sample of length $L=1.2\mu\text{m}$ (see inset in Fig.~4 of Ref.~\cite{Li15}). Matching with our theory of
many impurities leads to $2(1-K)\approx0.37$, or $K\approx0.82$.
Thus, in presence of many impurities, even moderately weak interactions
can give rise to the power law seen in Ref.~\cite{Li15}. The two
possible explanations (many impurities and weak interaction {\it vs}. single
impurity and strong interaction) of the observed conductance predict
different dependencies of $G$ on the edge length: for many impurities
one expects $N\propto L$ and hence resistive behavior $G\propto L^{-1}$.
Although $G(L)$ dependence is not reported in Ref.~\cite{Li15}, the earlier work~\cite{Du2015} found it to be linear at $L\gtrsim 10\mu{\rm m}$
~\footnote{Recently, in the topologically trivial regime (but still
edge-dominated) $G\propto L^{-1}$ has been observed even at sub-micron
lengths~\cite{Nichele2015}.}. The presence of magnetic impurities may also be identified from their subtle effect on 
the non-linear $I$-$V$ characteristics, which we discuss next.

\begin{figure} \includegraphics[trim={0cm 0cm 0cm 0cm},clip,width=\columnwidth]{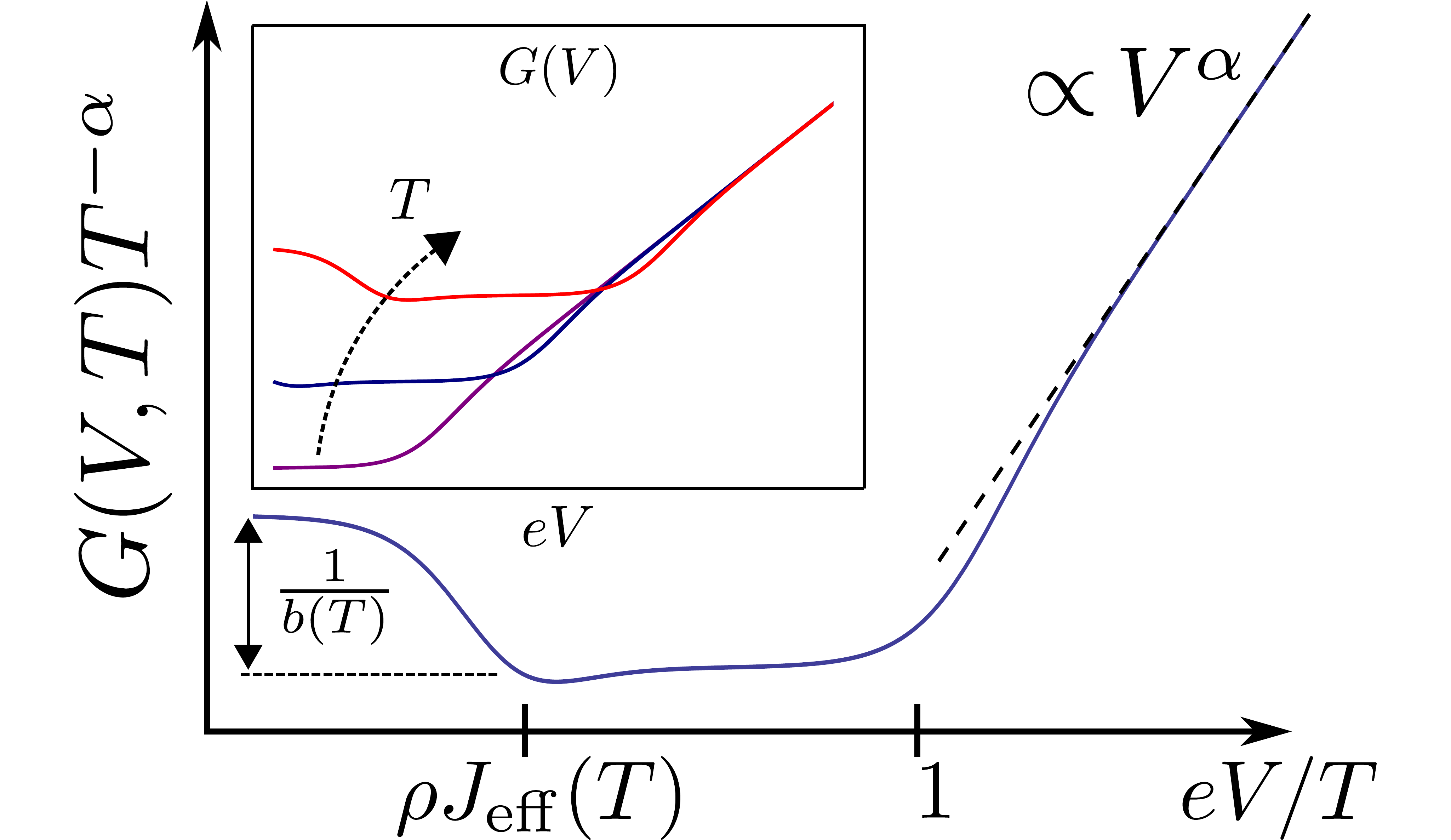} \caption{ (Color online) Log-log plot of the scaled conductance, $G\cdot T^{-\alpha}$,  in the presence of many impurities, $G \propto 1/ \delta G $. Here $\delta G = d \left\langle \delta I\right\rangle /dV$ and $\alpha = 2-2K>0$ are taken from Eq.~(\ref{eq:current3}), valid at intermediate energies $\max (T,eV)>T^*$.  The conductance has two crossover scales in its $V$-dependence. The higher crossover is at $eV \sim T$: above it, conductance increases (upon increasing $V$) asymptotically as a power law with exponent $\alpha >0$ (dashed line). Below it, $G$ stays roughly constant until the lower crossover scale, $eV \sim \rho J_{\text{eff}}(T) T$, is reached. Below it, the conductance changes by a factor $1/b(T)>1$ that depends weakly on temperature, see discussion below Eq.~(\ref{eq:C}). The inset shows $G(V)$ at three different temperatures $T$ (increasing from the lowest to highest curve).} \label{fig:1} \end{figure}

{\it Refinement of Eq.~(\ref{eq:QualitCurrentFinal}) -} 
The simplified form Eq.~\eqref{eq:QualitCurrentFinal} of the current-voltage characteristic misses several fine points relevant for the future analysis of experiments: (1) it does not provide the accurate form of the crossover at $eV/T\sim 1$, and (2)
it does not reveal an additional crossover at smaller bias, $eV/T\sim \rho J$.
The latter crossover is associated with the precession of the local magnetic moment in the exchange field $h\sim eV\rho J$ produced by the spins of itinerant edge electrons under a finite bias~\cite{Note2}. The crossover occurs once the precession frequency $\propto h$ becomes comparable to the Korringa relaxation rate,~\cite{korringa50} $1/\tau_K\sim (\rho J)^2 T$, as we will see in a detailed derivation of backscattering current.

The current operator of backscattered electrons is given by~\cite{KaneQH} $\delta I=-e\partial_{t}\delta N$
where $2\delta N=(N_{L}-N_{R})$ is the difference between the number
of left and right movers on the edge; it obeys $[\delta N,\,s_{i}(x_{0})]=i\epsilon_{zin}s_{n}(x_{0})$
and commutes with $H_{0}$. The decomposition~\eqref{eq:HamiltonianFull} of the Hamiltonian
is useful because at zero frequency the Hamiltonian $H_{\text{iso}}$,
Eq.~\eqref{eq:HamiltonianIso}, does not lead to backscattering of
helical edge electrons~\cite{tanaka_conductance_2011}.
It can be seen by noticing that: (i) $\partial_{t}\left\langle S_{z}\right\rangle =0$
in a steady state, because $S_{z}$ is bounded; this allows one to write the average backscattering current as~\cite{vayrynen_resistance_2014} $\left\langle \delta I\right\rangle =-e\partial_{t}\left\langle S_{z}^{tot}\right\rangle $
with $S_{z}^{tot}=\delta N+S_{z}$, and (ii) the operator $S_{z}^{tot}$
commutes with $H_{\text{iso}}$ and therefore is a conserved quantity
in absence of $\delta J_{ij}$. 
Hence $\partial_{t}\left\langle S_{z}^{tot}\right\rangle |_{\delta J\to0}=0$ and $\left\langle \delta I\right\rangle |_{\delta J\to0}=0$. 
We focus here on the case of a single magnetic moment; in the presence of many moments, we can define $S_{z}^{tot}=\delta N+\sum_{n} S^{(n)}_{z}$ where the sum is over the localized spins $\mathbf{S}^{(n)}$. 
In this work, we ignore the effects of correlations between the localized spins and coherent backscattering, allowing us to simply add up single-moment contributions to the edge resistance.  
This is justified for dilute spins, as discussed in more detail in Ref.~\cite{Note3}.

From hereon, we consider scattering off a single spin, and express the average steady-state backscattered
current as 
$\left\langle \delta I\right\rangle =-e\partial_{t}\left\langle S_{z}^{tot}\right\rangle $.
Commuting with the Hamiltonian~\eqref{eq:HamiltonianFull} leads
to [we denote $\delta J_{++}=\delta J_{xx}-\delta J_{yy}+i{(\delta J_{xy}+\delta J_{yx})},\,S_{\pm}=S_x \pm iS_y$ for brevity]
\begin{equation}
\begin{aligned}
 & \left\langle \delta I\right\rangle  =e\sum_{i,j=x,y} \epsilon_{ijz} \left(\delta J_{jz}\left\langle S_{i}:s_{z}:\right\rangle +\delta J_{zj}\left\langle S_{z}:s_{i}:\right\rangle \right) \\
& \!+e\text{Im}\delta J_{++} \left\langle S_{-}:\!s_{-}:\right\rangle +\! \frac{1}{2}\rho e^{2}V \left(\delta J_{yz}\left\langle S_{x}\right\rangle -\delta J_{xz}\left\langle S_{y}\right\rangle \right)\,. \\
\end{aligned}
\label{eq:current0}
\end{equation}
In agreement with the presence of an integral of motion, the average
current vanishes 
 when $\delta J\to0$.
The averaging above is done with respect to the density matrix $\varrho$ with Hamiltonian~\eqref{eq:HamiltonianFull}
in presence of a finite bias voltage, $\varrho\sim e^{-\beta\left(H-eVS_{z}^{tot}\right)}$~\cite{vayrynen_resistance_2014}.
We denote $:\! s_{j} \!: =s_{j}-\langle s_{j}\rangle_{0}$
with $\langle\rangle_{0}$ being the thermal average in absence of exchange
interaction, $\varrho_{0}\sim e^{-\beta\left(H_{0}-eV \delta N \right)}$.
The last term in \eqref{eq:current0} comes from the reducible part
$\langle s_{j}\rangle_{0}=\frac{1}{2}\delta_{jz}\rho eV$.
 
Equation~\eqref{eq:current0} is evaluated at time $t$ long enough
so that the steady-state value of $\left\langle \mathbf{S}\right\rangle $
has been reached. The averages $\langle S_{k}\!:\!s_{l}\!:\rangle$ can
be evaluated approximately in the exchange interaction assuming a
separation of time scales for the itinerant electron and spin dynamics~\cite{Note3}. The approximation results in 
\begin{equation}
\begin{aligned}\langle S_{k}:\negthinspace s_{l}\negthinspace:\rangle(t) & \approx-\sum_{j}(\delta_{kj}J_{0}+\delta J_{kj})\frac{1}{2}\text{Im}C_{jl}\\
 & -\sum_{ijn}(\delta_{ij}J_{0}+\delta J_{ij})\epsilon_{ikn}\left\langle S_{n}\right\rangle \text{Re}C_{jl}\,.
\end{aligned}
\label{eq:Sscorrelator}
\end{equation}
Here $\left\langle S_{n}\right\rangle $ is the steady-state impurity
spin polarization created by the current passing on the edge. The
integrated correlation function $C_{nl}=\int_{0}^{\infty}dt'\left\langle s_{n}(0)\!:\! s_{l}(t')\!:\right\rangle _{0}$
depends on temperature and bias voltage (through the average $\langle\dots\rangle_{0}$).
The only non-zero components of the matrix of $C_{nl}$ are the diagonals
and $C_{xy}\!=\!-C_{yx}\neq0$, the latter being due to finite bias voltage.
The temperature and bias dependence of $C_{nl}$ appearing in Eq.~\eqref{eq:Sscorrelator}
can be moved into the $T$ and $V$ dependence of running couplings $J_{ij}(T,V)$~\cite{Note3}. 
Inserting Eq.~\eqref{eq:Sscorrelator} into Eq.~\eqref{eq:current0} allows us to express the backscattering current in terms of the running couplings and steady-state values of the local-moment spin polarization $\left\langle \mathbf{S}\right\rangle $, see Ref.~\cite{Note3}. 
The last is found from the Bloch equations~\cite{bloch46}.
At $\delta J=0$, its only finite component is $\left\langle S_{z}\right\rangle =\frac{1}{2}\tanh\frac{eV}{2T}$
due to the $U(1)$ symmetry. Aiming at the lowest-order in $\delta J$ result for $\langle \delta I\rangle$, we need to find $\left\langle S_{x,y}\right\rangle$ to the first order in $\delta J$. Unlike
$\left\langle S_{z}\right\rangle $, which is a function of $eV/T$
given by thermodynamics, the components $\left\langle S_{x,y}\right\rangle $ depend~\cite{Note3} on both the effective field $h_{z}=\frac{1}{2}eV\rho J_{z}$
generated by the bias voltage, and on the local-moment Korringa relaxation rate $\tau_{K}^{-1}=\frac{\pi}{2}\rho^{2}(J_{\perp}^{2}\frac{\frac{eV}{2T}}{\tanh\frac{eV}{2T}}+J_{z}^{2})T$. (We use here the running couplings with their implicit dependence on $V$ and $T$.) 
The backscattering current is 
\begin{equation}
\begin{aligned}\left\langle \delta I\right\rangle  & =e\frac{\pi }{4}eV\rho^{2}|\delta J_{++}(T,V)|^{2} +e\frac{\pi }{4}eV\frac{1}{2}R(T,V) \\
 & \times \sum_{i=x,y}\rho^{2}\left(\delta J_{zi}(T,V)+\frac{J_{\perp}(T,V)}{J_{z}(T,V)}\delta J_{iz}(T,V)\right)^{\!2}\!.
\end{aligned} 
\label{eq:currentGeneralRunningCouplings}
\end{equation}
Here the first term arises from non-zero $\left\langle S_{z}\right\rangle $
and can be derived simply from Fermi Golden Rule by assuming $\left\langle \mathbf{S}\right\rangle =\mathbf{z}\frac{1}{2}\tanh\frac{eV}{2T}$.
In the second term, function 
\begin{equation}
R(T,V)\approx\frac{\frac{J_{z}(T,0)}{J_{\text{eff}}(T)}+x^{2}}{1+x^{2}}\,,\quad
x=\frac{eV}{2T}\frac{2/\pi}{ \rho J_{\text{eff}}(T)}\,,
\label{eq:R}
\end{equation}
comes from $\left\langle S_{x,y}\right\rangle\neq 0$ and therefore depends on the ratio $h_{z}/\tau_{K} = x$. 
Here we abbreviated $\rho J_{\text{eff}}(T)=\rho {[J_{\perp}(T,0)^{2}+J_{z}(T,0)^{2}]}/J_{z}(T,0) \ll 1 $. 
In Eq.~(\ref{eq:R}) the term $J_{z}/J_{\text{eff}}\lesssim1$ only matters at very small bias $eV\ll T\rho J_{\text{eff}}\ll T$; thus we have neglected the $V$-dependence in it. 


In Eq.~\eqref{eq:currentGeneralRunningCouplings} the current is written
in terms of the running couplings $J_{ij}(T,\,V)$. Next, we will
write it in terms of the bare couplings, which allows us to see explicitly
the $T,V$-dependence of $\left\langle \delta I\right\rangle $. At
$T^{*}<\max(eV,\,T)<D$ one has~\cite{Note3} 
\begin{equation}
X(T,V)\approx X(D)\left( \frac{D}{2\pi T} \right)^{1-K}\sqrt{F(\frac{eV}{2T})}\label{eq:JperpConj}
\end{equation}
with a function 
\begin{equation}
F(y)=KB(K+i\frac{y}{\pi},K-i\frac{y}{\pi})\frac{\sinh y}{y}\approx\frac{B(K,K)}{[1+A(K)y^{2}]^{1-K}}\,.
\label{eq:F}
\end{equation}
Here $A(K)=\pi^{-2}\Gamma(K)^{\frac{2}{1-K}}$ and $B$ is the Euler Beta function; $X$ stands for
any of the quantities, $\text{Re}\delta J_{++},\,\text{Im}\delta J_{++},$ and $\delta J_{zi}+\frac{J_{\perp}}{J_{z}}\delta J_{iz}$
($i=x,\,y$), which appear in Eq.~\eqref{eq:currentGeneralRunningCouplings}.

Using Eqs.~\eqref{eq:currentGeneralRunningCouplings}--\eqref{eq:F}
 we arrive at the central result of this paper: the temperature and bias dependence of the current can be lumped in a product of several simple terms,
\begin{flalign}
&\left\langle \delta I\right\rangle =\delta G_0 
\left[\frac{D}{2\pi T}\right]^{2-2K}V\frac{B(K,K)}{[1+A(K)(\frac{eV}{2T})^{2}]^{1-K}} f(x,T)\,,\nonumber\\
 &f(x,T)=\frac{b(T)+x^{2}}{1+x^{2}}\,,\,\,x=\frac{eV}{2T}\frac{2/\pi}{\rho J_{\text{eff}}(T)}. \label{eq:current3}
\end{flalign}
Here
  the $T$-independent factor is $\delta G_0=\frac{e^{2}}{\hbar}\frac{\pi}{4}\rho^{2}\delta J_{\text{tot}}^{2}(D)\,,$
\begin{equation}
\delta J_{\text{tot}}^{2}(D)\negthinspace=|\delta J_{++}(D)|^{2}+\frac{1}{2}\sum_{i=x,y}[\delta J_{zi}(D)+\delta J_{iz}(D)]^{2}\,,\label{eq:E}
\end{equation}
while $J_{\text{eff}}(T)$ and 
\begin{equation}
b(T)\negthinspace=1-\frac{(1-\frac{J_{z}(T,0)}{J_{\text{eff}}(T)})\frac{1}{2}\sum_{i=x,y}[\delta J_{zi}(D)+\delta J_{iz}(D)]^{2}}{|\delta J_{++}(D)|^{2}+\frac{1}{2}\sum_{i=x,y}[\delta J_{zi}(D)+\delta J_{iz}(D)]^{2}}\,\label{eq:C}
\end{equation}
display a weak temperature dependence~\cite{Note3}.
(For typical values of exchange couplings $\delta J_{ij}(D)$ 
function $b(T)$ can be well approximated
by a constant of order~1: $0.67\leq b(T)\leq0.83$ in the interval
$T^{*}\leq T\leq D$~\cite{Note3}.)  At a fixed temperature $T$,
the current dependence on bias $V$ has two well-separated crossover scales described by the last two factors in~\eqref{eq:current3}. The smaller scale, $V\sim T\rho J_{\text{eff}}(T)$, is associated with the impurity spin dynamics. 
The crossover at the higher scale, $V\sim T$, occurs between the linear and weakly-nonlinear $\langle\delta I\rangle$ vs. $V$ dependencies. Near this crossover one may set $f\to1$ in Eq.~\eqref{eq:current3}, reproducing the
result of Eq.~\eqref{eq:QualitCurrentFinal} with, however, accurate crossover behavior near $eV\sim T$.

{\it The backscattering current at energies below $T^*$ -} 
At energies $E\lesssim T^{*}$, one may neglect the small term $\propto (1-K)$ in~\eqref{eq:RGJperp}--\eqref{eq:RGJz} and consider the resulting weak-coupling Kondo RG with the initial condition $\rho J_{\perp}(T^{*})=\sqrt{2}(1-K)$~\cite{Note3}. For small $1-K$, it yields the Kondo temperature $T_{K}\sim T^{*}e^{-1/\sqrt{2}(1-K)}\ll T^*$. The RG flow erases the uniaxial anisotropy created by $K \neq 1$, and $J_z \approx J_{\perp} $ at energies below $T^*$. As a result, $J_{\text{eff}}=2 J_z$ in Eq.~(\ref{eq:R}) and $R=(\frac{1}{2}+x^2)/(1+x^2)$. 
Similarly, the anisotropic perturbation in Eq.~(\ref{eq:HamiltonianFull}) becomes RG-irrelevant, and Eq.~(\ref{eq:JperpConj}) is replaced~\cite{vayrynen_resistance_2014} by $X(E) \approx X(T^*) \frac{\ln E/T_{K}}{\ln T^*/T_{K}}$. Hence, the backscattering current becomes
\begin{equation}
\left\langle \delta I\right\rangle = \delta G_0 V \left[\frac{\ln\max(T,eV)/T_{K}}{\ln T^*/T_{K}}\right]^{2} \frac{b+x^{2}}{1+x^{2}}\,,\label{eq:currentbelowTstar}
\end{equation}
valid for $T_K < \max(T,eV) < T^*$. Here $b$ is given by Eq.~(\ref{eq:C}) which becomes independent of $T$ upon setting $J_{\text{eff}}=2 J_z$. Similarly, 
 $\delta G_0$ was introduced below Eq.~(\ref{eq:current3}) but now one must use $\delta J_{\text{tot}}^{2}(T^*)$  in it with the ``new'' bare cutoff. 

The coupling constant $\rho J_z(E)\sim [\ln(E/T_K)]^{-1}$ grows in the course of RG, and below the Kondo temperature, $\max(T,eV)<T_{K}$,  Eqs.~\eqref{eq:RGJperp}--\eqref{eq:RGJz} are no longer valid. In this regime one can use the phenomenological local-interaction Hamiltonian~\cite{schmidt_inelastic_2012,lezmy_single_2012} 
 to obtain $\delta G(V,T)\propto T^4 g(V/T)$; the crossover function $g(x)$ has asymptotes $g(x\to 0)=\rm const$ and $g(x\gg 1)\sim x^4$. Details can be found in Ref.~\cite{lezmy_single_2012} upon setting $K=1$ therein.
  Note that $\delta G$ decreases when reducing $T,\,eV$ and thus leads to $G=e^2/h$ in the limit of zero temperature and bias. This behavior is opposite from Eq.~(\ref{eq:current3}) which indicated an insulating edge at low energies. 

{\it Conclusions -} 
We analyzed the joint effect of two weak interactions on the edge conduction in a 2D topological insulator. These interactions are: the repulsion between itinerant electrons of an edge state, and their exchange with the local magnetic moments. This joint effect may result in a seemingly insulating behavior of the edge conduction down to a low temperature scale $T^*$, see Eq.~(\ref{eq:TstarDef}): at $\max(T,eV)\gtrsim T^*$, the single-impurity backscattering current $\left\langle \delta I\right\rangle$ \textit{grows} as a power law upon lowering temperature  or bias, see Eq.~\eqref{eq:current3}, or Fig.~\ref{fig:1} for the conductance in presence of many moments. Localized magnetic moments may appear in a narrow-gap semiconductor as  a consequence of charge disorder~\cite{vayrynen_resistance_2014}. 
Scattering off magnetic moments provides an alternative explanation of the recent experiment~\cite{Li15}, assuming $T^*$ is below the temperature range explored in~\cite{Li15}. 
 [None of the considered interactions break the time-reversal symmetry~\footnote{we disregard here the possibility of spontaneous symmetry breaking~\cite{Altshuler13,Yudson15}.}, so at low energies, $\max(T,eV) \ll T^*$, backscattering is suppressed, see Eq.~(\ref{eq:currentbelowTstar}).]
 The developed theory is also applicable to magnetically-doped~\cite{Jungwirth06}~\footnote{We expect $T^{*} \sim 0.9 \mathrm{mK}$ for a Mn-doped InAs-based heterostructure~\cite{Wang14}. To arrive at this estimate we used the following parameters: exchange coupling $J/a_{0}^{3} \sim 1 eV$ per unit cell volume~\cite{Wang14}, lattice constant $a_0 \approx 0.6 \mathrm{nm}$, quantum well thickness $d\approx 11.5 \mathrm{nm}$, edge state velocity $v\approx 5.7\times 10^{4} \mathrm{m/s}$, penetration depth $\xi \approx 16 \mathrm{nm}$, $E_g \approx 54 \mathrm{K}$~\cite{Li15}, and $K \approx 0.8$.} 
 heterostructures. 
 Finally, we find two crossovers in the $I$-$V$ characteristics: the main one occurs at $eV\sim T$; 
 a more subtle one occurs at lower bias, $eV \sim \rho J T$, see Fig.~\ref{fig:1}. Its observation in future experiments may provide evidence for the considered mechanism of the edge state excess resistance.

\begin{acknowledgments}
We thank Richard Brierley, Rui-Rui Du, and Hendrik Meier for discussions. This work was supported by NSF DMR Grant No. 1206612 and DFG through SFB 1170 "ToCoTronics".
\end{acknowledgments}

\bibliographystyle{apsrev4-1}
\bibliography{refs}

\begin{thebibliography}{41}%
\makeatletter
\providecommand \@ifxundefined [1]{%
 \@ifx{#1\undefined}
}%
\providecommand \@ifnum [1]{%
 \ifnum #1\expandafter \@firstoftwo
 \else \expandafter \@secondoftwo
 \fi
}%
\providecommand \@ifx [1]{%
 \ifx #1\expandafter \@firstoftwo
 \else \expandafter \@secondoftwo
 \fi
}%
\providecommand \natexlab [1]{#1}%
\providecommand \enquote  [1]{``#1''}%
\providecommand \bibnamefont  [1]{#1}%
\providecommand \bibfnamefont [1]{#1}%
\providecommand \citenamefont [1]{#1}%
\providecommand \href@noop [0]{\@secondoftwo}%
\providecommand \href [0]{\begingroup \@sanitize@url \@href}%
\providecommand \@href[1]{\@@startlink{#1}\@@href}%
\providecommand \@@href[1]{\endgroup#1\@@endlink}%
\providecommand \@sanitize@url [0]{\catcode `\\12\catcode `\$12\catcode
  `\&12\catcode `\#12\catcode `\^12\catcode `\_12\catcode `\%12\relax}%
\providecommand \@@startlink[1]{}%
\providecommand \@@endlink[0]{}%
\providecommand \url  [0]{\begingroup\@sanitize@url \@url }%
\providecommand \@url [1]{\endgroup\@href {#1}{\urlprefix }}%
\providecommand \urlprefix  [0]{URL }%
\providecommand \Eprint [0]{\href }%
\providecommand \doibase [0]{http://dx.doi.org/}%
\providecommand \selectlanguage [0]{\@gobble}%
\providecommand \bibinfo  [0]{\@secondoftwo}%
\providecommand \bibfield  [0]{\@secondoftwo}%
\providecommand \translation [1]{[#1]}%
\providecommand \BibitemOpen [0]{}%
\providecommand \bibitemStop [0]{}%
\providecommand \bibitemNoStop [0]{.\EOS\space}%
\providecommand \EOS [0]{\spacefactor3000\relax}%
\providecommand \BibitemShut  [1]{\csname bibitem#1\endcsname}%
\let\auto@bib@innerbib\@empty
\bibitem [{\citenamefont {Li}\ \emph {et~al.}(2015)\citenamefont {Li},
  \citenamefont {Wang}, \citenamefont {Fu}, \citenamefont {Du}, \citenamefont
  {Schreiber}, \citenamefont {Mu}, \citenamefont {Liu}, \citenamefont
  {Sullivan}, \citenamefont {Cs\'athy}, \citenamefont {Lin},\ and\
  \citenamefont {Du}}]{Li15}%
  \BibitemOpen
  \bibfield  {author} {\bibinfo {author} {\bibfnamefont {T.}~\bibnamefont
  {Li}}, \bibinfo {author} {\bibfnamefont {P.}~\bibnamefont {Wang}}, \bibinfo
  {author} {\bibfnamefont {H.}~\bibnamefont {Fu}}, \bibinfo {author}
  {\bibfnamefont {L.}~\bibnamefont {Du}}, \bibinfo {author} {\bibfnamefont
  {K.~A.}\ \bibnamefont {Schreiber}}, \bibinfo {author} {\bibfnamefont
  {X.}~\bibnamefont {Mu}}, \bibinfo {author} {\bibfnamefont {X.}~\bibnamefont
  {Liu}}, \bibinfo {author} {\bibfnamefont {G.}~\bibnamefont {Sullivan}},
  \bibinfo {author} {\bibfnamefont {G.~A.}\ \bibnamefont {Cs\'athy}}, \bibinfo
  {author} {\bibfnamefont {X.}~\bibnamefont {Lin}}, \ and\ \bibinfo {author}
  {\bibfnamefont {R.-R.}\ \bibnamefont {Du}},\ }\href {\doibase
  10.1103/PhysRevLett.115.136804} {\bibfield  {journal} {\bibinfo  {journal}
  {Phys. Rev. Lett.}\ }\textbf {\bibinfo {volume} {115}},\ \bibinfo {pages}
  {136804} (\bibinfo {year} {2015})}\BibitemShut {NoStop}%
\bibitem [{\citenamefont {Liu}\ \emph {et~al.}(2008)\citenamefont {Liu},
  \citenamefont {Hughes}, \citenamefont {Qi}, \citenamefont {Wang},\ and\
  \citenamefont {Zhang}}]{liu08}%
  \BibitemOpen
  \bibfield  {author} {\bibinfo {author} {\bibfnamefont {C.}~\bibnamefont
  {Liu}}, \bibinfo {author} {\bibfnamefont {T.~L.}\ \bibnamefont {Hughes}},
  \bibinfo {author} {\bibfnamefont {X.-L.}\ \bibnamefont {Qi}}, \bibinfo
  {author} {\bibfnamefont {K.}~\bibnamefont {Wang}}, \ and\ \bibinfo {author}
  {\bibfnamefont {S.-C.}\ \bibnamefont {Zhang}},\ }\href {\doibase
  10.1103/PhysRevLett.100.236601} {\bibfield  {journal} {\bibinfo  {journal}
  {Phys. Rev. Lett.}\ }\textbf {\bibinfo {volume} {100}},\ \bibinfo {pages}
  {236601} (\bibinfo {year} {2008})}\BibitemShut {NoStop}%
\bibitem [{\citenamefont {Knez}\ \emph {et~al.}(2014)\citenamefont {Knez},
  \citenamefont {Rettner}, \citenamefont {Yang}, \citenamefont {Parkin},
  \citenamefont {Du}, \citenamefont {Du},\ and\ \citenamefont
  {Sullivan}}]{Knez14}%
  \BibitemOpen
  \bibfield  {author} {\bibinfo {author} {\bibfnamefont {I.}~\bibnamefont
  {Knez}}, \bibinfo {author} {\bibfnamefont {C.~T.}\ \bibnamefont {Rettner}},
  \bibinfo {author} {\bibfnamefont {S.-H.}\ \bibnamefont {Yang}}, \bibinfo
  {author} {\bibfnamefont {S.~S.~P.}\ \bibnamefont {Parkin}}, \bibinfo {author}
  {\bibfnamefont {L.}~\bibnamefont {Du}}, \bibinfo {author} {\bibfnamefont
  {R.-R.}\ \bibnamefont {Du}}, \ and\ \bibinfo {author} {\bibfnamefont
  {G.}~\bibnamefont {Sullivan}},\ }\href {\doibase
  10.1103/PhysRevLett.112.026602} {\bibfield  {journal} {\bibinfo  {journal}
  {Phys. Rev. Lett.}\ }\textbf {\bibinfo {volume} {112}},\ \bibinfo {pages}
  {026602} (\bibinfo {year} {2014})}\BibitemShut {NoStop}%
\bibitem [{\citenamefont {Spanton}\ \emph {et~al.}(2014)\citenamefont
  {Spanton}, \citenamefont {Nowack}, \citenamefont {Du}, \citenamefont
  {Sullivan}, \citenamefont {Du},\ and\ \citenamefont {Moler}}]{Spanton04}%
  \BibitemOpen
  \bibfield  {author} {\bibinfo {author} {\bibfnamefont {E.~M.}\ \bibnamefont
  {Spanton}}, \bibinfo {author} {\bibfnamefont {K.~C.}\ \bibnamefont {Nowack}},
  \bibinfo {author} {\bibfnamefont {L.}~\bibnamefont {Du}}, \bibinfo {author}
  {\bibfnamefont {G.}~\bibnamefont {Sullivan}}, \bibinfo {author}
  {\bibfnamefont {R.-R.}\ \bibnamefont {Du}}, \ and\ \bibinfo {author}
  {\bibfnamefont {K.~A.}\ \bibnamefont {Moler}},\ }\href {\doibase
  10.1103/PhysRevLett.113.026804} {\bibfield  {journal} {\bibinfo  {journal}
  {Phys. Rev. Lett.}\ }\textbf {\bibinfo {volume} {113}},\ \bibinfo {pages}
  {026804} (\bibinfo {year} {2014})}\BibitemShut {NoStop}%
\bibitem [{\citenamefont {Du}\ \emph {et~al.}(2015)\citenamefont {Du},
  \citenamefont {Knez}, \citenamefont {Sullivan},\ and\ \citenamefont
  {Du}}]{Du2015}%
  \BibitemOpen
  \bibfield  {author} {\bibinfo {author} {\bibfnamefont {L.}~\bibnamefont
  {Du}}, \bibinfo {author} {\bibfnamefont {I.}~\bibnamefont {Knez}}, \bibinfo
  {author} {\bibfnamefont {G.}~\bibnamefont {Sullivan}}, \ and\ \bibinfo
  {author} {\bibfnamefont {R.-R.}\ \bibnamefont {Du}},\ }\href {\doibase
  10.1103/PhysRevLett.114.096802} {\bibfield  {journal} {\bibinfo  {journal}
  {Phys. Rev. Lett.}\ }\textbf {\bibinfo {volume} {114}},\ \bibinfo {pages}
  {096802} (\bibinfo {year} {2015})}\BibitemShut {NoStop}%
\bibitem [{\citenamefont {{Nichele}}\ \emph {et~al.}(2015)\citenamefont
  {{Nichele}}, \citenamefont {{Suominen}}, \citenamefont {{Kjaergaard}},
  \citenamefont {{Marcus}}, \citenamefont {{Sajadi}}, \citenamefont {{Folk}},
  \citenamefont {{Qu}}, \citenamefont {{Beukman}}, \citenamefont {{de Vries}},
  \citenamefont {{van Veen}}, \citenamefont {{Nadj-Perge}}, \citenamefont
  {{Kouwenhoven}}, \citenamefont {{Nguyen}}, \citenamefont {{Kiselev}},
  \citenamefont {{Yi}}, \citenamefont {{Sokolich}}, \citenamefont {{Manfra}},
  \citenamefont {{Spanton}},\ and\ \citenamefont {{Moler}}}]{Nichele2015}%
  \BibitemOpen
  \bibfield  {author} {\bibinfo {author} {\bibfnamefont {F.}~\bibnamefont
  {{Nichele}}}, \bibinfo {author} {\bibfnamefont {H.~J.}\ \bibnamefont
  {{Suominen}}}, \bibinfo {author} {\bibfnamefont {M.}~\bibnamefont
  {{Kjaergaard}}}, \bibinfo {author} {\bibfnamefont {C.~M.}\ \bibnamefont
  {{Marcus}}}, \bibinfo {author} {\bibfnamefont {E.}~\bibnamefont {{Sajadi}}},
  \bibinfo {author} {\bibfnamefont {J.~A.}\ \bibnamefont {{Folk}}}, \bibinfo
  {author} {\bibfnamefont {F.}~\bibnamefont {{Qu}}}, \bibinfo {author}
  {\bibfnamefont {A.~J.~A.}\ \bibnamefont {{Beukman}}}, \bibinfo {author}
  {\bibfnamefont {F.~K.}\ \bibnamefont {{de Vries}}}, \bibinfo {author}
  {\bibfnamefont {J.}~\bibnamefont {{van Veen}}}, \bibinfo {author}
  {\bibfnamefont {S.}~\bibnamefont {{Nadj-Perge}}}, \bibinfo {author}
  {\bibfnamefont {L.~P.}\ \bibnamefont {{Kouwenhoven}}}, \bibinfo {author}
  {\bibfnamefont {B.-M.}\ \bibnamefont {{Nguyen}}}, \bibinfo {author}
  {\bibfnamefont {A.~A.}\ \bibnamefont {{Kiselev}}}, \bibinfo {author}
  {\bibfnamefont {W.}~\bibnamefont {{Yi}}}, \bibinfo {author} {\bibfnamefont
  {M.}~\bibnamefont {{Sokolich}}}, \bibinfo {author} {\bibfnamefont {M.~J.}\
  \bibnamefont {{Manfra}}}, \bibinfo {author} {\bibfnamefont {E.~M.}\
  \bibnamefont {{Spanton}}}, \ and\ \bibinfo {author} {\bibfnamefont {K.~A.}\
  \bibnamefont {{Moler}}},\ }\href@noop {} {\bibfield  {journal} {\bibinfo
  {journal} {ArXiv e-prints}\ } (\bibinfo {year} {2015})},\ \Eprint
  {http://arxiv.org/abs/1511.01728} {arXiv:1511.01728 [cond-mat.mes-hall]}
  \BibitemShut {NoStop}%
\bibitem [{\citenamefont {Kane}\ and\ \citenamefont {Fisher}(1992)}]{KFPRL92}%
  \BibitemOpen
  \bibfield  {author} {\bibinfo {author} {\bibfnamefont {C.~L.}\ \bibnamefont
  {Kane}}\ and\ \bibinfo {author} {\bibfnamefont {M.~P.~A.}\ \bibnamefont
  {Fisher}},\ }\href {\doibase 10.1103/PhysRevLett.68.1220} {\bibfield
  {journal} {\bibinfo  {journal} {Phys. Rev. Lett.}\ }\textbf {\bibinfo
  {volume} {68}},\ \bibinfo {pages} {1220} (\bibinfo {year}
  {1992})}\BibitemShut {NoStop}%
\bibitem [{\citenamefont {Wu}\ \emph {et~al.}(2006)\citenamefont {Wu},
  \citenamefont {Bernevig},\ and\ \citenamefont {Zhang}}]{wu_helical_2006}%
  \BibitemOpen
  \bibfield  {author} {\bibinfo {author} {\bibfnamefont {C.}~\bibnamefont
  {Wu}}, \bibinfo {author} {\bibfnamefont {B.~A.}\ \bibnamefont {Bernevig}}, \
  and\ \bibinfo {author} {\bibfnamefont {S.-C.}\ \bibnamefont {Zhang}},\ }\href
  {\doibase 10.1103/PhysRevLett.96.106401} {\bibfield  {journal} {\bibinfo
  {journal} {Phys. Rev. Lett.}\ }\textbf {\bibinfo {volume} {96}},\ \bibinfo
  {pages} {106401} (\bibinfo {year} {2006})}\BibitemShut {NoStop}%
\bibitem [{\citenamefont {Xu}\ and\ \citenamefont {Moore}(2006)}]{Xu06}%
  \BibitemOpen
  \bibfield  {author} {\bibinfo {author} {\bibfnamefont {C.}~\bibnamefont
  {Xu}}\ and\ \bibinfo {author} {\bibfnamefont {J.~E.}\ \bibnamefont {Moore}},\
  }\href {\doibase 10.1103/PhysRevB.73.045322} {\bibfield  {journal} {\bibinfo
  {journal} {Phys. Rev. B}\ }\textbf {\bibinfo {volume} {73}},\ \bibinfo
  {pages} {045322} (\bibinfo {year} {2006})}\BibitemShut {NoStop}%
\bibitem [{\citenamefont {V\"ayrynen}\ \emph {et~al.}(2013)\citenamefont
  {V\"ayrynen}, \citenamefont {Goldstein},\ and\ \citenamefont
  {Glazman}}]{vayrynen_helical_2013}%
  \BibitemOpen
  \bibfield  {author} {\bibinfo {author} {\bibfnamefont {J.~I.}\ \bibnamefont
  {V\"ayrynen}}, \bibinfo {author} {\bibfnamefont {M.}~\bibnamefont
  {Goldstein}}, \ and\ \bibinfo {author} {\bibfnamefont {L.~I.}\ \bibnamefont
  {Glazman}},\ }\href {\doibase 10.1103/PhysRevLett.110.216402} {\bibfield
  {journal} {\bibinfo  {journal} {Phys. Rev. Lett.}\ }\textbf {\bibinfo
  {volume} {110}},\ \bibinfo {pages} {216402} (\bibinfo {year}
  {2013})}\BibitemShut {NoStop}%
\bibitem [{\citenamefont {V\"ayrynen}\ \emph {et~al.}(2014)\citenamefont
  {V\"ayrynen}, \citenamefont {Goldstein}, \citenamefont {Gefen},\ and\
  \citenamefont {Glazman}}]{vayrynen_resistance_2014}%
  \BibitemOpen
  \bibfield  {author} {\bibinfo {author} {\bibfnamefont {J.~I.}\ \bibnamefont
  {V\"ayrynen}}, \bibinfo {author} {\bibfnamefont {M.}~\bibnamefont
  {Goldstein}}, \bibinfo {author} {\bibfnamefont {Y.}~\bibnamefont {Gefen}}, \
  and\ \bibinfo {author} {\bibfnamefont {L.~I.}\ \bibnamefont {Glazman}},\
  }\href {\doibase 10.1103/PhysRevB.90.115309} {\bibfield  {journal} {\bibinfo
  {journal} {Phys. Rev. B}\ }\textbf {\bibinfo {volume} {90}},\ \bibinfo
  {pages} {115309} (\bibinfo {year} {2014})}\BibitemShut {NoStop}%
\bibitem [{\citenamefont {S\'en\'echal}(2004)}]{senechal}%
  \BibitemOpen
  \bibfield  {author} {\bibinfo {author} {\bibfnamefont {D.}~\bibnamefont
  {S\'en\'echal}},\ }in\ \href {\doibase 10.1007/0-387-21717-7_4} {\emph
  {\bibinfo {booktitle} {Theoretical Methods for Strongly Correlated
  Electrons}}},\ \bibinfo {series and number} {CRM Series in Mathematical
  Physics},\ \bibinfo {editor} {edited by\ \bibinfo {editor} {\bibfnamefont
  {D.}~\bibnamefont {S\'en\'echal}}, \bibinfo {editor} {\bibfnamefont {A.-M.}\
  \bibnamefont {Tremblay}}, \ and\ \bibinfo {editor} {\bibfnamefont
  {C.}~\bibnamefont {Bourbonnais}}}\ (\bibinfo  {publisher} {Springer New
  York},\ \bibinfo {year} {2004})\ pp.\ \bibinfo {pages} {139--186}\BibitemShut
  {NoStop}%
\bibitem [{\citenamefont {Lee}\ and\ \citenamefont
  {Toner}(1992)}]{lee_kondo_1992}%
  \BibitemOpen
  \bibfield  {author} {\bibinfo {author} {\bibfnamefont {D.-H.}\ \bibnamefont
  {Lee}}\ and\ \bibinfo {author} {\bibfnamefont {J.}~\bibnamefont {Toner}},\
  }\href {\doibase 10.1103/PhysRevLett.69.3378} {\bibfield  {journal} {\bibinfo
   {journal} {Phys. Rev. Lett.}\ }\textbf {\bibinfo {volume} {69}},\ \bibinfo
  {pages} {3378} (\bibinfo {year} {1992})}\BibitemShut {NoStop}%
\bibitem [{\citenamefont {Furusaki}\ and\ \citenamefont
  {Nagaosa}(1994)}]{furusaki_kondo_1994}%
  \BibitemOpen
  \bibfield  {author} {\bibinfo {author} {\bibfnamefont {A.}~\bibnamefont
  {Furusaki}}\ and\ \bibinfo {author} {\bibfnamefont {N.}~\bibnamefont
  {Nagaosa}},\ }\href {\doibase 10.1103/PhysRevLett.72.892} {\bibfield
  {journal} {\bibinfo  {journal} {Phys. Rev. Lett.}\ }\textbf {\bibinfo
  {volume} {72}},\ \bibinfo {pages} {892} (\bibinfo {year} {1994})}\BibitemShut
  {NoStop}%
\bibitem [{\citenamefont {Maciejko}\ \emph {et~al.}(2009)\citenamefont
  {Maciejko}, \citenamefont {Liu}, \citenamefont {Oreg}, \citenamefont {Qi},
  \citenamefont {Wu},\ and\ \citenamefont {Zhang}}]{Maciejko2009}%
  \BibitemOpen
  \bibfield  {author} {\bibinfo {author} {\bibfnamefont {J.}~\bibnamefont
  {Maciejko}}, \bibinfo {author} {\bibfnamefont {C.}~\bibnamefont {Liu}},
  \bibinfo {author} {\bibfnamefont {Y.}~\bibnamefont {Oreg}}, \bibinfo {author}
  {\bibfnamefont {X.-L.}\ \bibnamefont {Qi}}, \bibinfo {author} {\bibfnamefont
  {C.}~\bibnamefont {Wu}}, \ and\ \bibinfo {author} {\bibfnamefont {S.-C.}\
  \bibnamefont {Zhang}},\ }\href {\doibase 10.1103/PhysRevLett.102.256803}
  {\bibfield  {journal} {\bibinfo  {journal} {Phys. Rev. Lett.}\ }\textbf
  {\bibinfo {volume} {102}},\ \bibinfo {pages} {256803} (\bibinfo {year}
  {2009})}\BibitemShut {NoStop}%
\bibitem [{\citenamefont {Cardy}(1996)}]{cardy1996scaling}%
  \BibitemOpen
  \bibfield  {author} {\bibinfo {author} {\bibfnamefont {J.}~\bibnamefont
  {Cardy}},\ }\href@noop {} {\emph {\bibinfo {title} {Scaling and
  renormalization in statistical physics}}},\ Vol.~\bibinfo {volume} {5}\
  (\bibinfo  {publisher} {Cambridge university press},\ \bibinfo {year}
  {1996})\BibitemShut {NoStop}%
\bibitem [{\citenamefont {{Anderson}}(1970)}]{anderson_poor_1970}%
  \BibitemOpen
  \bibfield  {author} {\bibinfo {author} {\bibfnamefont {P.}~\bibnamefont
  {{Anderson}}},\ }\href@noop {} {\bibfield  {journal} {\bibinfo  {journal}
  {Journal of {Physics} {C}: {Solid} {State} {Physics}}\ }\textbf {\bibinfo
  {volume} {3}},\ \bibinfo {pages} {2436} (\bibinfo {year} {1970})}\BibitemShut
  {NoStop}%
\bibitem [{\citenamefont {Giamarchi}(2003)}]{Giamarchi}%
  \BibitemOpen
  \bibfield  {author} {\bibinfo {author} {\bibfnamefont {T.}~\bibnamefont
  {Giamarchi}},\ }\href@noop {} {\emph {\bibinfo {title} {Quantum Physics in
  One Dimension}}},\ International Series of Monographs on Physics\ (\bibinfo
  {publisher} {Clarendon Press},\ \bibinfo {year} {2003})\BibitemShut {NoStop}%
\bibitem [{Note1()}]{Note1}%
  \BibitemOpen
  \bibinfo {note} {In general, the cutoff depends on the microscopic structure
  of the impurity. If the exchange term in Eq.~\protect \textup {\hbox
  {\mathsurround \z@ \protect \normalfont (\ignorespaces \ref
  {eq:HamiltonianFull}\unskip \@@italiccorr )}} originates from a charge
  puddle, the cutoff is $D\sim \protect \qopname \relax m{min}(E_{g},\protect
  \tmspace +\thinmuskip {.1667em}E_{C},\protect \tmspace +\thinmuskip
  {.1667em}\delta )$, where $E_C$ and $\delta $ are, respectively, the energies
  of charged and chargeless excitations in a puddle~\cite
  {vayrynen_resistance_2014}.}\BibitemShut {Stop}%
\bibitem [{\citenamefont {Tanaka}\ \emph {et~al.}(2011)\citenamefont {Tanaka},
  \citenamefont {Furusaki},\ and\ \citenamefont
  {Matveev}}]{tanaka_conductance_2011}%
  \BibitemOpen
  \bibfield  {author} {\bibinfo {author} {\bibfnamefont {Y.}~\bibnamefont
  {Tanaka}}, \bibinfo {author} {\bibfnamefont {A.}~\bibnamefont {Furusaki}}, \
  and\ \bibinfo {author} {\bibfnamefont {K.~A.}\ \bibnamefont {Matveev}},\
  }\href {\doibase 10.1103/PhysRevLett.106.236402} {\bibfield  {journal}
  {\bibinfo  {journal} {Phys. Rev. Lett.}\ }\textbf {\bibinfo {volume} {106}},\
  \bibinfo {pages} {236402} (\bibinfo {year} {2011})}\BibitemShut {NoStop}%
\bibitem [{Note2()}]{Note2}%
  \BibitemOpen
  \bibinfo {note} {The bias voltage creates an imbalance between left and right
  movers and a non-zero $ \langle s_{z} \rangle $ in Eq.~(\ref
  {eq:HamiltonianIso}). This results in non-zero $ \langle S_{z} \rangle $
  through the exchange interaction~(\ref {eq:HamiltonianIso})}\BibitemShut
  {NoStop}%
\bibitem [{Note3()}]{Note3}%
  \BibitemOpen
  \bibinfo {note} {See Supplemental Material for details. The supplement
  includes references to~\protect \rev@citealp
  {BreuerPetruccione,Gradshteyn,GiamarchiSchulz,Gornyi05,Dyakonov,maciejko_kondo_2012,eriksson_spin-orbit_2013}}\BibitemShut
  {NoStop}%
\bibitem [{Note4()}]{Note4}%
  \BibitemOpen
  \bibinfo {note} {Recently, in the topologically trivial regime (but still
  edge-dominated) $G\propto L^{-1}$ has been observed even at sub-micron
  lengths~\cite {Nichele2015}.}\BibitemShut {Stop}%
\bibitem [{\citenamefont {Korringa}(1950)}]{korringa50}%
  \BibitemOpen
  \bibfield  {author} {\bibinfo {author} {\bibfnamefont {J.}~\bibnamefont
  {Korringa}},\ }\href {\doibase
  http://dx.doi.org/10.1016/0031-8914(50)90105-4} {\bibfield  {journal}
  {\bibinfo  {journal} {Physica}\ }\textbf {\bibinfo {volume} {16}},\ \bibinfo
  {pages} {601 } (\bibinfo {year} {1950})}\BibitemShut {NoStop}%
\bibitem [{\citenamefont {Kane}\ and\ \citenamefont {Fisher}(1994)}]{KaneQH}%
  \BibitemOpen
  \bibfield  {author} {\bibinfo {author} {\bibfnamefont {C.~L.}\ \bibnamefont
  {Kane}}\ and\ \bibinfo {author} {\bibfnamefont {M.~P.~A.}\ \bibnamefont
  {Fisher}},\ }\href {\doibase 10.1103/PhysRevLett.72.724} {\bibfield
  {journal} {\bibinfo  {journal} {Phys. Rev. Lett.}\ }\textbf {\bibinfo
  {volume} {72}},\ \bibinfo {pages} {724} (\bibinfo {year} {1994})}\BibitemShut
  {NoStop}%
\bibitem [{\citenamefont {Bloch}(1946)}]{bloch46}%
  \BibitemOpen
  \bibfield  {author} {\bibinfo {author} {\bibfnamefont {F.}~\bibnamefont
  {Bloch}},\ }\href {\doibase 10.1103/PhysRev.70.460} {\bibfield  {journal}
  {\bibinfo  {journal} {Phys. Rev.}\ }\textbf {\bibinfo {volume} {70}},\
  \bibinfo {pages} {460} (\bibinfo {year} {1946})}\BibitemShut {NoStop}%
\bibitem [{\citenamefont {Schmidt}\ \emph {et~al.}(2012)\citenamefont
  {Schmidt}, \citenamefont {Rachel}, \citenamefont {von Oppen},\ and\
  \citenamefont {Glazman}}]{schmidt_inelastic_2012}%
  \BibitemOpen
  \bibfield  {author} {\bibinfo {author} {\bibfnamefont {T.~L.}\ \bibnamefont
  {Schmidt}}, \bibinfo {author} {\bibfnamefont {S.}~\bibnamefont {Rachel}},
  \bibinfo {author} {\bibfnamefont {F.}~\bibnamefont {von Oppen}}, \ and\
  \bibinfo {author} {\bibfnamefont {L.~I.}\ \bibnamefont {Glazman}},\ }\href
  {\doibase 10.1103/PhysRevLett.108.156402} {\bibfield  {journal} {\bibinfo
  {journal} {Phys. Rev. Lett.}\ }\textbf {\bibinfo {volume} {108}},\ \bibinfo
  {pages} {156402} (\bibinfo {year} {2012})}\BibitemShut {NoStop}%
\bibitem [{\citenamefont {Lezmy}\ \emph {et~al.}(2012)\citenamefont {Lezmy},
  \citenamefont {Oreg},\ and\ \citenamefont {Berkooz}}]{lezmy_single_2012}%
  \BibitemOpen
  \bibfield  {author} {\bibinfo {author} {\bibfnamefont {N.}~\bibnamefont
  {Lezmy}}, \bibinfo {author} {\bibfnamefont {Y.}~\bibnamefont {Oreg}}, \ and\
  \bibinfo {author} {\bibfnamefont {M.}~\bibnamefont {Berkooz}},\ }\href
  {\doibase 10.1103/PhysRevB.85.235304} {\bibfield  {journal} {\bibinfo
  {journal} {Phys. Rev. B}\ }\textbf {\bibinfo {volume} {85}},\ \bibinfo
  {pages} {235304} (\bibinfo {year} {2012})}\BibitemShut {NoStop}%
\bibitem [{Note5()}]{Note5}%
  \BibitemOpen
  \bibinfo {note} {We disregard here the possibility of spontaneous symmetry
  breaking~\cite {Altshuler13,Yudson15}.}\BibitemShut {Stop}%
\bibitem [{\citenamefont {Jungwirth}\ \emph {et~al.}(2006)\citenamefont
  {Jungwirth}, \citenamefont {Sinova}, \citenamefont {Ma\ifmmode~\check{s}\else
  \v{s}\fi{}ek}, \citenamefont {Ku\ifmmode~\check{c}\else \v{c}\fi{}era},\ and\
  \citenamefont {MacDonald}}]{Jungwirth06}%
  \BibitemOpen
  \bibfield  {author} {\bibinfo {author} {\bibfnamefont {T.}~\bibnamefont
  {Jungwirth}}, \bibinfo {author} {\bibfnamefont {J.}~\bibnamefont {Sinova}},
  \bibinfo {author} {\bibfnamefont {J.}~\bibnamefont {Ma\ifmmode~\check{s}\else
  \v{s}\fi{}ek}}, \bibinfo {author} {\bibfnamefont {J.}~\bibnamefont
  {Ku\ifmmode~\check{c}\else \v{c}\fi{}era}}, \ and\ \bibinfo {author}
  {\bibfnamefont {A.~H.}\ \bibnamefont {MacDonald}},\ }\href {\doibase
  10.1103/RevModPhys.78.809} {\bibfield  {journal} {\bibinfo  {journal} {Rev.
  Mod. Phys.}\ }\textbf {\bibinfo {volume} {78}},\ \bibinfo {pages} {809}
  (\bibinfo {year} {2006})}\BibitemShut {NoStop}%
\bibitem [{Note6()}]{Note6}%
  \BibitemOpen
  \bibinfo {note} {We expect $T^{*} \sim 0.9 \protect \mathrm {mK}$ for a
  Mn-doped InAs-based heterostructure~\cite {Wang14}. To arrive at this
  estimate we used the following parameters: exchange coupling $J/a_{0}^{3}
  \sim 1 eV$ per unit cell volume~\cite {Wang14}, lattice constant $a_0 \approx
  0.6 \protect \mathrm {nm}$, quantum well thickness $d\approx 11.5 \protect
  \mathrm {nm}$, edge state velocity $v\approx 5.7\times 10^{4} \protect
  \mathrm {m/s}$, penetration depth $\xi \approx 16 \protect \mathrm {nm}$,
  $E_g \approx 54 \protect \mathrm {K}$~\cite {Li15}, and $K \approx
  0.8$.}\BibitemShut {Stop}%
\bibitem [{\citenamefont {Breuer}\ and\ \citenamefont
  {Petruccione}(2002)}]{BreuerPetruccione}%
  \BibitemOpen
  \bibfield  {author} {\bibinfo {author} {\bibfnamefont {H.-P.}\ \bibnamefont
  {Breuer}}\ and\ \bibinfo {author} {\bibfnamefont {F.}~\bibnamefont
  {Petruccione}},\ }\href@noop {} {\emph {\bibinfo {title} {The theory of open
  quantum systems}}}\ (\bibinfo  {publisher} {Oxford University Press on
  Demand},\ \bibinfo {year} {2002})\BibitemShut {NoStop}%
\bibitem [{\citenamefont {Gradshteyn}\ \emph {et~al.}(1994)\citenamefont
  {Gradshteyn}, \citenamefont {Ryzhik},\ and\ \citenamefont
  {Jeffrey}}]{Gradshteyn}%
  \BibitemOpen
  \bibfield  {author} {\bibinfo {author} {\bibfnamefont {I.}~\bibnamefont
  {Gradshteyn}}, \bibinfo {author} {\bibfnamefont {I.}~\bibnamefont {Ryzhik}},
  \ and\ \bibinfo {author} {\bibfnamefont {A.}~\bibnamefont {Jeffrey}},\
  }\href@noop {} {\emph {\bibinfo {title} {Table of Integrals, Series and
  Products 5th edn (New York: Academic)}}}\ (\bibinfo {year}
  {1994})\BibitemShut {NoStop}%
\bibitem [{\citenamefont {Giamarchi}\ and\ \citenamefont
  {Schulz}(1988)}]{GiamarchiSchulz}%
  \BibitemOpen
  \bibfield  {author} {\bibinfo {author} {\bibfnamefont {T.}~\bibnamefont
  {Giamarchi}}\ and\ \bibinfo {author} {\bibfnamefont {H.~J.}\ \bibnamefont
  {Schulz}},\ }\href {\doibase 10.1103/PhysRevB.37.325} {\bibfield  {journal}
  {\bibinfo  {journal} {Phys. Rev. B}\ }\textbf {\bibinfo {volume} {37}},\
  \bibinfo {pages} {325} (\bibinfo {year} {1988})}\BibitemShut {NoStop}%
\bibitem [{\citenamefont {Gornyi}\ \emph {et~al.}(2005)\citenamefont {Gornyi},
  \citenamefont {Mirlin},\ and\ \citenamefont {Polyakov}}]{Gornyi05}%
  \BibitemOpen
  \bibfield  {author} {\bibinfo {author} {\bibfnamefont {I.~V.}\ \bibnamefont
  {Gornyi}}, \bibinfo {author} {\bibfnamefont {A.~D.}\ \bibnamefont {Mirlin}},
  \ and\ \bibinfo {author} {\bibfnamefont {D.~G.}\ \bibnamefont {Polyakov}},\
  }\href {\doibase 10.1103/PhysRevLett.95.046404} {\bibfield  {journal}
  {\bibinfo  {journal} {Phys. Rev. Lett.}\ }\textbf {\bibinfo {volume} {95}},\
  \bibinfo {pages} {046404} (\bibinfo {year} {2005})}\BibitemShut {NoStop}%
\bibitem [{\citenamefont {Dyakonov}(1994)}]{Dyakonov}%
  \BibitemOpen
  \bibfield  {author} {\bibinfo {author} {\bibfnamefont {M.}~\bibnamefont
  {Dyakonov}},\ }\href {\doibase
  http://dx.doi.org/10.1016/0038-1098(94)90459-6} {\bibfield  {journal}
  {\bibinfo  {journal} {Solid State Communications}\ }\textbf {\bibinfo
  {volume} {92}},\ \bibinfo {pages} {711 } (\bibinfo {year}
  {1994})}\BibitemShut {NoStop}%
\bibitem [{\citenamefont {Maciejko}(2012)}]{maciejko_kondo_2012}%
  \BibitemOpen
  \bibfield  {author} {\bibinfo {author} {\bibfnamefont {J.}~\bibnamefont
  {Maciejko}},\ }\href {\doibase 10.1103/PhysRevB.85.245108} {\bibfield
  {journal} {\bibinfo  {journal} {Phys. Rev. B}\ }\textbf {\bibinfo {volume}
  {85}},\ \bibinfo {pages} {245108} (\bibinfo {year} {2012})}\BibitemShut
  {NoStop}%
\bibitem [{\citenamefont {Eriksson}(2013)}]{eriksson_spin-orbit_2013}%
  \BibitemOpen
  \bibfield  {author} {\bibinfo {author} {\bibfnamefont {E.}~\bibnamefont
  {Eriksson}},\ }\href {\doibase 10.1103/PhysRevB.87.235414} {\bibfield
  {journal} {\bibinfo  {journal} {Phys. Rev. B}\ }\textbf {\bibinfo {volume}
  {87}},\ \bibinfo {pages} {235414} (\bibinfo {year} {2013})}\BibitemShut
  {NoStop}%
\bibitem [{\citenamefont {Altshuler}\ \emph {et~al.}(2013)\citenamefont
  {Altshuler}, \citenamefont {Aleiner},\ and\ \citenamefont
  {Yudson}}]{Altshuler13}%
  \BibitemOpen
  \bibfield  {author} {\bibinfo {author} {\bibfnamefont {B.~L.}\ \bibnamefont
  {Altshuler}}, \bibinfo {author} {\bibfnamefont {I.~L.}\ \bibnamefont
  {Aleiner}}, \ and\ \bibinfo {author} {\bibfnamefont {V.~I.}\ \bibnamefont
  {Yudson}},\ }\href {\doibase 10.1103/PhysRevLett.111.086401} {\bibfield
  {journal} {\bibinfo  {journal} {Phys. Rev. Lett.}\ }\textbf {\bibinfo
  {volume} {111}},\ \bibinfo {pages} {086401} (\bibinfo {year}
  {2013})}\BibitemShut {NoStop}%
\bibitem [{\citenamefont {Yevtushenko}\ \emph {et~al.}(2015)\citenamefont
  {Yevtushenko}, \citenamefont {Wugalter}, \citenamefont {Yudson},\ and\
  \citenamefont {Altshuler}}]{Yudson15}%
  \BibitemOpen
  \bibfield  {author} {\bibinfo {author} {\bibfnamefont {O.~M.}\ \bibnamefont
  {Yevtushenko}}, \bibinfo {author} {\bibfnamefont {A.}~\bibnamefont
  {Wugalter}}, \bibinfo {author} {\bibfnamefont {V.~I.}\ \bibnamefont
  {Yudson}}, \ and\ \bibinfo {author} {\bibfnamefont {B.~L.}\ \bibnamefont
  {Altshuler}},\ }\href {http://stacks.iop.org/0295-5075/112/i=5/a=57003}
  {\bibfield  {journal} {\bibinfo  {journal} {EPL (Europhysics Letters)}\
  }\textbf {\bibinfo {volume} {112}},\ \bibinfo {pages} {57003} (\bibinfo
  {year} {2015})}\BibitemShut {NoStop}%
\bibitem [{\citenamefont {Wang}\ \emph {et~al.}(2014)\citenamefont {Wang},
  \citenamefont {Liu}, \citenamefont {Zhang}, \citenamefont {Samarth},
  \citenamefont {Zhang},\ and\ \citenamefont {Liu}}]{Wang14}%
  \BibitemOpen
  \bibfield  {author} {\bibinfo {author} {\bibfnamefont {Q.-Z.}\ \bibnamefont
  {Wang}}, \bibinfo {author} {\bibfnamefont {X.}~\bibnamefont {Liu}}, \bibinfo
  {author} {\bibfnamefont {H.-J.}\ \bibnamefont {Zhang}}, \bibinfo {author}
  {\bibfnamefont {N.}~\bibnamefont {Samarth}}, \bibinfo {author} {\bibfnamefont
  {S.-C.}\ \bibnamefont {Zhang}}, \ and\ \bibinfo {author} {\bibfnamefont
  {C.-X.}\ \bibnamefont {Liu}},\ }\href {\doibase
  10.1103/PhysRevLett.113.147201} {\bibfield  {journal} {\bibinfo  {journal}
  {Phys. Rev. Lett.}\ }\textbf {\bibinfo {volume} {113}},\ \bibinfo {pages}
  {147201} (\bibinfo {year} {2014})}\BibitemShut {NoStop}%
\end{thebibliography}%



\section*{Supplementary material}

# Supplemental Material to "Magnetic moments in a helical edge can make weak correlations seem strong"

Jukka I. Väyrynen,[1] Florian Geissler,[2] and Leonid I. Glazman[1]

[1] *Department of Physics, Yale University, New Haven, CT 06520, USA*
[2] *Institute for Theoretical Physics and Astrophysics,
University of Würzburg, 97074 Würzburg, Germany*

In this supplemental material we show in detail the derivations that were omitted in the main text.

## SM1. DERIVATION OF THE EQ. (9) OF THE MAIN TEXT

In this section, we express the long time averages $\langle S_k(t) \!:\! s_l(t) \!:\rangle$ in terms of the steady-state values $\langle \mathbf{S}(t) \rangle$ of the impurity spin. As in the main text, we denote $: s_j := s_j - \langle s_j \rangle_0$. While the unperturbed average $\langle S_k(t) \!:\! s_l(t) \!:\rangle_0$ factorizes and vanishes (since $\langle : s_l(t) : \rangle_0 = 0$), the average $\langle S_k(t) \!:\! s_l(t) \!:\rangle$ in the presence of the exchange coupling is in general finite and starts from first order in $J$. In this section, we consider a generic exchange Hamiltonian $\sum_{ij} J_{ij} S_i s_j$ without distinguishing isotropic and anisotropic parts of $J_{ij}$. We mostly discuss the case of a single local moment; in Sec. SM1 A we explain how the results, and Eqs. (8)-(9) of the main text, are modified if there are many moments, and discuss the conditions to ignore these modifications, as was done in the main text.

In the interaction picture, $A^I(t) = e^{iH_0 t} A e^{-iH_0 t}$, the local spin remains time-independent $S_k^I(t) = S_k$ because the isolated edge Hamiltonian $H_0$ commutes with $\mathbf{S}$. The average of an observable $A$ can be written as $\langle A(t) \rangle = \mathrm{Tr} \varrho(t) A = \mathrm{Tr} \varrho^I(t) A^I(t)$. The Heisenberg equation of motion for the density matrix is in integral form $\varrho^I(t) = \varrho_i \otimes \varrho_0 + i \int_{-\infty}^{t} dt' [\varrho^I(t'), \sum_{ij} J_{ij} S_i s_j^I(t')]$. Here $\varrho_i$ and $\varrho_0 \sim e^{-\beta(H_0 - eV\delta N)}$ are the respective density matrices of the decoupled impurity spin and helical edge system far away in the past (we imagine an adiabatic turn-on of the exchange interaction). Using the expression for $\varrho^I(t)$, we find

$$\langle S_k(t) \!:\! s_l(t) \!:\rangle = i \int_{-\infty}^{t} dt' \mathrm{Tr} \varrho(t') e^{-iH_0 t'} \Big[ \sum_{ij} J_{ij} S_i s_j^I(t'), S_k \!:\! s_l^I(t) \!:\Big] e^{iH_0 t'}$$

$$= i \sum_{ij} J_{ij} \int_{-\infty}^{t} dt' \mathrm{Tr} \varrho(t') e^{-iH_0 t'} \left( \frac{1}{2} i \sum_n \epsilon_{ikn} S_n \{ s_j^I(t'), \!:\! s_l^I(t) \!:\} + \frac{1}{4} \delta_{ik} [s_j^I(t'), \!:\! s_l^I(t) \!:] \right) e^{iH_0 t'}. \quad (S1)$$

We used $\{S_i, S_k\} = \frac{1}{2}\delta_{ik}$. Next, we approximate $\varrho(t') \approx \varrho_S(t) \otimes \varrho_0$ which is valid since the $s$-$s$ correlation functions of itinerant electrons decay much faster than $\langle \mathbf{S}(t) \rangle$ changes [S1] (see also the end of this section). The approximation leads to

$$\langle S_k(t) \!:\! s_l(t) \!:\rangle \approx i \sum_{ij} J_{ij} \int_{-\infty}^{t} dt' \left( \frac{1}{2} i \sum_n \epsilon_{ikn} \langle S_n(t) \rangle \langle \{ s_j^I(t'), \!:\! s_l^I(t) \!:\} \rangle_0 + \frac{1}{4} \delta_{ik} \langle [s_j^I(t'), \!:\! s_l^I(t) \!:] \rangle_0 \right) \quad (S2)$$

$$\approx -\sum_{ij} J_{ij} \left( \sum_n \epsilon_{ikn} \langle S_n(t) \rangle \mathrm{Re} C_{jl} + \frac{1}{2} \delta_{ik} \mathrm{Im} C_{jl} \right), \quad (S3)$$

with $C_{jl} = \int_0^\infty dt' \langle s_j^I(0) \!:\! s_l^I(t') \!:\rangle_0$. This is Eq. (9) of the main text. Above $\langle S_n(t) \rangle$ is the steady-state polarization of the local moment. In the next section we derive an equation of motion for it with the help of Eq. (S3).

The spin density-density correlation functions in $C_{jl}$ can be calculated using standard bosonization techniques [S2]. The bias voltage dependence can be moved into the time evolution; for example,

$$\langle s_y^I(t) \!:\! s_x \!:\rangle_0 = -\langle s_x^I(t) \!:\! s_y^I \!:\rangle_0 = \langle s_x^I(t) \!:\! s_x^I \!:\rangle_{0,V\to 0} \sin eVt\,; \quad \langle s_x^I(t) \!:\! s_x^I \!:\rangle_0 = \langle s_y^I(t) \!:\! s_y^I \!:\rangle_0 = \langle s_x^I(t) \!:\! s_x^I \!:\rangle_{0,V\to 0} \cos eVt\,, \quad (S4)$$

where the average $\langle \ldots \rangle_{0,V\to 0}$ is evaluated without bias voltage: $\langle s_x^I(t) \!:\! s_x^I \!:\rangle_{0,V\to 0} \propto \sinh^{-2K} \pi T(t - i\frac{a}{v})$. Here $a$ is the short distance cutoff (in the main text, we pick $a = v/D$ for convenience). In addition to the terms in Eq. (S4), only $\langle s_z(t) \!:\! s_z \!:\rangle_0 = \langle s_z(t) \!:\! s_z \!:\rangle_{0,V\to 0}$ is non-vanishing (and it is independent of bias voltage).

We need the following integrated correlation functions, [S2, S3]

$$\mathrm{Re}C_{xx} = \frac{K^2\pi}{2}T\rho^2\left(\frac{v/a}{2\pi T}\right)^{2-2K} B\left(K+i\frac{eV}{2\pi T}, K-i\frac{eV}{2\pi T}\right)\cosh\left(\frac{eV}{2T}\right), \tag{S5}$$

$$\mathrm{Re}C_{zz} = \frac{K\pi}{2}T\rho^2, \tag{S6}$$

$$\mathrm{Re}C_{xy} = \frac{K^2\pi}{2}T\rho^2\left(\frac{v/a}{2\pi T}\right)^{2-2K} B\left(K+i\frac{eV}{2\pi T}, K-i\frac{eV}{2\pi T}\right)\sinh\left(\frac{eV}{2T}\right)\cot\pi K, \tag{S7}$$

$$\mathrm{Im}C_{xy} = \frac{K^2\pi}{2}T\rho^2\left(\frac{v/a}{2\pi T}\right)^{2-2K} B\left(K+i\frac{eV}{2\pi T}, K-i\frac{eV}{2\pi T}\right)\sinh\left(\frac{eV}{2T}\right), \tag{S8}$$

while $C_{yy} = C_{xx}$, $C_{yx} = -C_{xy}$. Above $B(x,y)$ is the Euler Beta function. We introduced density of states $\rho = 1/(2\pi vK)$. In the main text we approximate $K \approx 1$ in the prefactors. In Eq. (S7) we have neglected a correction of order $(\frac{E}{v/a})^{2-2K}$, where $E \sim \max(T, eV) \ll v/a$. In the limit $K \to 1$ this correction becomes important and will cancel the divergence from $\cot \pi K$; one then finds $\mathrm{Re}C_{xy} \sim eV \ln E$.

Before discussing the case of many local moments, let us briefly investigate the conditions for the validity of the approximation $\langle S_n(t')\rangle \approx \langle S_n(t)\rangle$ in Eq. (S2). Without the approximation we would have $\int_0^\infty dt' \langle S_n(t-t')\rangle \langle s_j^I(0): s_l^I(t'):\rangle_0$ in the first term of Eq. (S3). The approximation $\langle S_n(t-t')\rangle \approx \langle S_n(t)\rangle$ inside the integrand is justified if the shortest time scale of variation of $\langle S_n \rangle$ is large compared to that of $\langle s_j^I(0): s_l^I(t'):\rangle_0$. The latter varies on the time scale $\min(1/eV, 1/T)$, see Eq. (S4) and the expression for $\langle s_x^I(t): s_x^I:\rangle_{0,V\to 0}$ below it. The time scale for $\langle S_n \rangle$ is given by $\min[1/\rho JeV, 1/(\rho J)^2 T]$. It is obtained from the Bloch equation (S11) below, where the dynamics of the impurity spin is seen to have two time scales: the period of precession $1/h \sim 1/\rho JeV$ and the relaxation time $1/\gamma \sim 1/[(\rho J)^2 \max(T, eV)]$. It is easy to check that at weak coupling, $\rho J \ll 1$, we always have $\min[1/\rho JeV, 1/(\rho J)^2 T] \gg \min(1/eV, 1/T)$ which justifies the approximation in Eq. (S2).

### A. Multiple local moments

Finally, let us comment on how the electron interference effects may affect the above results and Eqs. (8)-(9) of the main text. For that purpose, we retain in the exchange Hamiltonian a sum over all the local moments: $\sum_p \sum_{ij} J_{ij}^{(p)} S_i^{(p)} s_j(x_p)$ at positions $\{x_p\}$. As a result, Eqs. (8)-(9) of the main text become modified. In Eq. (8) the modification is simple: the relevant quantities will get labels, $J, S, s \to J^{(p)}, S^{(p)}, s(x_p)$, and there will be a sum over $p$ (coming from the commutator of $S_z^{tot}$ with the exchange Hamiltonian). Modification of Eq. (S3) [Eq. (9) of the main text] is less simple as it contains contributions from coupling different spins at $x_p$ and $x_q$:

$$\langle S_k^{(p)}(t): s_l(x_p, t):\rangle \approx -\sum_{ij} J_{ij}^{(p)}\left(\sum_n \epsilon_{ikn}\langle S_n^{(p)}(t)\rangle \mathrm{Re}C_{jl} + \frac{1}{2}\delta_{ik}\mathrm{Im}C_{jl}\right)$$
$$-2\sum_{q\neq p}\sum_{ij} J_{ij}^{(q)}\langle S_i^{(q)}(t) S_k^{(p)}(t)\rangle \mathrm{Im}C_{jl}^{qp}, \tag{S9}$$

where the first line is what we had in Eq. (S3). The new term is the second line which will give interference contributions to the backscattering current (from spins at $x_p$ and $x_q$); in it we introduced the correlation function that couples the two sites, $C_{jl}^{qp} = \int_0^\infty dt' \langle s_j^I(0,0): s_l^I(x_p - x_q, t'):\rangle_0$. For a low density of moments (see Refs. [S4, S5]), we can ignore collective effects and approximate $\langle S_i^{(q)}(t) S_k^{(p)}(t)\rangle = \langle S_i^{(q)}(t)\rangle \langle S_k^{(p)}(t)\rangle$ in Eq. (S9).

At small bias $eV \ll T$ all the spins are only weakly polarized $\langle S_i^{(q)}(t)\rangle \sim eV/T$, and the second line of Eq. (S9) is smaller than the first line by a factor $\sim (eV/T)^2$. Hence, at low bias coherent backscattering off two spins gives a negligible correction to the backscattering current $\langle \delta I\rangle \sim eV$.

At very high bias, $eV \gg T$, the passing current on the edge almost fully polarizes the impurity spins (see Ref. [21] of the main text). We then have $\langle S_i^{(q)}(t)\rangle \approx \delta_{iz}/2$ and the second line of Eq. (S9) becomes comparable to the first one. In principle, this additional term contributes to the backscattering current. Consider, for example, Eq. (8) of the main text which gives the average backscattering current from impurity spin at position $x_0$. In Eq. (8), the term $\delta J_{zy}^{(0)} \langle S_z^{(0)}: s_x(x_0):\rangle$ obtains an additional contribution [from using Eq. (S9)] $\sim \delta J_{zy}^{(0)} \delta J_{zx}^{(1)}$ when a second impurity spin $\mathbf{S}^{(1)}$ is present. Such terms in the backscattering current correspond to interference effects that arise from electron scattering off different localized spins. Note that these terms are possible because of the exchange couplings

$J_{zx}^{(p)}, J_{zy}^{(p)} \neq 0$ which backscatter the edge electron without flipping the local moments – interference can thus arise even when the magnetic moments are not correlated with each other. [The interference terms are also found by generalizing the Fermi Golden Rule result (7) of the main text to multiple impurities.]

The interference terms may lead to Anderson localization of the edge electrons [S6], unless dephasing destroys the effect [S7]. To demonstrate the suppression of the weak localization corrections in the above formalism, one needs to perform yet another iteration in the evaluation of the spin correlation function: that allows one to include scattering off three impurities [S8]. The dephasing rate $\tau_\phi^{-1}$ has been evaluated in Ref. [S7]. From [S7] we find $\tau_\phi^{-1} \sim (1-K)(En_i\rho J^2)^{1/2}$, where $E = \max(eV, T)$ is the electron energy, and $n_i$ is the local moment density. Localization corrections are small if $\tau_\phi^{-1}$ is much larger than the scattering rate $\tau^{-1} \sim n_i\rho J^2$ off local moments; this leads to the condition $\max(eV, T) > n_i\rho J^2/(1-K)^2$. Hence, for dilute enough local moments, we may ignore the interference corrections to current, as was done in the main text.

## SM2. THE BLOCH EQUATIONS

In this section, we derive the steady-state polarization $\langle \mathbf{S}(t) \rangle$ of the local moment. This can be done with the help of Eq. (S3) derived in the previous section. Writing the Heisenberg equation motion yields ($i = x, y, z$)

$$\frac{d}{dt}\langle S_i(t)\rangle = -\frac{1}{2}\rho eV \sum_{kn} J_{kz}\epsilon_{kin}\langle S_n(t)\rangle - \sum_{jkn} J_{kj}\epsilon_{kin}\langle S_n(t) \colon s_j(t) \colon \rangle. \tag{S10}$$

We used $\langle s_j(t)\rangle_0 = \frac{1}{2}\delta_{jz}\rho eV$ in the first term. Next, we use Eq. (S3) in the second term. The resulting equation can be cast in the form of a Bloch equation [S9]

$$\frac{d}{dt}\langle \mathbf{S}(t)\rangle = \mathbf{h} \times \langle \mathbf{S}(t)\rangle - \boldsymbol{\gamma}\langle \mathbf{S}\rangle + \mathbf{c}, \tag{S11}$$

where $\mathbf{h}$ is a torque created by the exchange field,

$$\mathbf{h} = \begin{pmatrix} \frac{1}{2}\rho eV J_{xz} - (J_{zy}J_{yx} - J_{zx}J_{yy})\mathrm{Re}C_{xy} \\ \frac{1}{2}\rho eV J_{yz} + (J_{zy}J_{xx} - J_{zx}J_{xy})\mathrm{Re}C_{xy} \\ \frac{1}{2}\rho eV J_{zz} - (J_{yy}J_{xx} - J_{yx}J_{xy})\mathrm{Re}C_{xy} \end{pmatrix}, \tag{S12}$$

$\boldsymbol{\gamma}$ is a symmetric tensor that describes the Korringa relaxation [S10] of the spin,

$$\boldsymbol{\gamma} = -\begin{pmatrix} -\sum_{k\neq x}\left((J_{kx}^2 + J_{ky}^2)\mathrm{Re}C_{xx} + J_{kz}^2\mathrm{Re}C_{zz}\right) & (J_{xx}J_{yx} + J_{xy}J_{yy})\mathrm{Re}C_{xx} + J_{xz}J_{yz}\mathrm{Re}C_{zz} & (J_{xx}J_{zx} + J_{xy}J_{zy})\mathrm{Re}C_{xx} + J_{xz}J_{zz}\mathrm{Re}C_{zz} \\ (J_{xx}J_{yx} + J_{xy}J_{yy})\mathrm{Re}C_{xx} + J_{xz}J_{yz}\mathrm{Re}C_{zz} & -\sum_{k\neq y}\left((J_{kx}^2 + J_{ky}^2)\mathrm{Re}C_{xx} + J_{kz}^2\mathrm{Re}C_{zz}\right) & (J_{yx}J_{zx} + J_{yy}J_{zy})\mathrm{Re}C_{xx} + J_{yz}J_{zz}\mathrm{Re}C_{zz} \\ (J_{xx}J_{zx} + J_{xy}J_{zy})\mathrm{Re}C_{xx} + J_{xz}J_{zz}\mathrm{Re}C_{zz} & (J_{yx}J_{zx} + J_{yy}J_{zy})\mathrm{Re}C_{xx} + J_{yz}J_{zz}\mathrm{Re}C_{zz} & -\sum_{k\neq z}\left((J_{kx}^2 + J_{ky}^2)\mathrm{Re}C_{xx} + J_{kz}^2\mathrm{Re}C_{zz}\right) \end{pmatrix}, \tag{S13}$$

and

$$\mathbf{c} = \begin{pmatrix} (J_{zy}J_{yx} - J_{yy}J_{zx})\mathrm{Im}C_{xy} \\ (J_{xy}J_{zx} - J_{zy}J_{xx})\mathrm{Im}C_{xy} \\ (J_{yy}J_{xx} - J_{xy}J_{yx})\mathrm{Im}C_{xy} \end{pmatrix}. \tag{S14}$$

The steady-state value $\langle \mathbf{S}(t)\rangle$ of the local-moment polarization can be expressed in terms of $\boldsymbol{\gamma}$, $\mathbf{h}$, and $\mathbf{c}$ by setting $d\langle \mathbf{S}(t)\rangle/dt = 0$ in Eq. (S11). Next, we shall do this to first order in $\delta J_{ij}$, assuming an almost isotropic exchange tensor, $J_{ij} = J_0\delta_{ij} + \delta J_{ij}$, with $\delta J_{ij} \ll J_0$. The quantities $\boldsymbol{\gamma}$, $\mathbf{h}$, and $\mathbf{c}$ can then be linearized in $\delta J$: $\boldsymbol{\gamma} \approx \boldsymbol{\gamma}^{(0)} + \boldsymbol{\gamma}^{(1)}$ etc. For illustration, we have

$$\boldsymbol{\gamma}^{(0)} = J_0^2 \begin{pmatrix} \mathrm{Re}C_{xx} + \mathrm{Re}C_{zz} & 0 & 0 \\ 0 & \mathrm{Re}C_{xx} + \mathrm{Re}C_{zz} & 0 \\ 0 & 0 & 2\mathrm{Re}C_{xx} \end{pmatrix}, \tag{S15}$$

$$\mathbf{h}^{(0)} = \mathbf{z}\left(\frac{1}{2}\rho J_0 eV - J_0^2 \mathrm{Re}C_{xy}\right), \quad \mathbf{h}^{(1)} = \begin{pmatrix} \frac{1}{2}\rho eV\delta J_{xz} + J_0\delta J_{zx}\mathrm{Re}C_{xy} \\ \frac{1}{2}\rho eV\delta J_{yz} + J_0\delta J_{zy}\mathrm{Re}C_{xy} \\ \frac{1}{2}\rho eV\delta J_{zz} - J_0(\delta J_{yy} + \delta J_{xx})\mathrm{Re}C_{xy} \end{pmatrix}. \tag{S16}$$

So far, all the exchange couplings are bare, $J_{ij} = J_{ij}(D)$. We will outline in Sec. SM4 A the procedure to express $\boldsymbol{\gamma}$, $\mathbf{h}$, and $\mathbf{c}$ in terms of running couplings. We find (for clarity we do not write explicitly the $T,V$-dependence of the couplings in $\boldsymbol{\gamma}$)

$$\boldsymbol{\gamma}^{(0)} = \begin{pmatrix} J_\perp^2 \frac{\frac{eV}{2T}}{\tanh\frac{eV}{2T}} + J_z^2 & 0 & 0 \\ 0 & J_\perp^2 \frac{\frac{eV}{2T}}{\tanh\frac{eV}{2T}} + J_z^2 & 0 \\ 0 & 0 & 2J_\perp^2 \frac{\frac{eV}{2T}}{\tanh\frac{eV}{2T}} \end{pmatrix} \mathrm{Re}C_{zz},$$

$$\boldsymbol{\gamma}^{(1)} = -\begin{pmatrix} -2\left(J_\perp \delta J_{yy}\frac{\frac{eV}{2T}}{\tanh\frac{eV}{2T}} + J_z\delta J_{zz}\right) & J_\perp(\delta J_{yx}+\delta J_{xy})\frac{\frac{eV}{2T}}{\tanh\frac{eV}{2T}} & J_\perp \delta J_{zx}\frac{\frac{eV}{2T}}{\tanh\frac{eV}{2T}} + J_z\delta J_{xz} \\ J_\perp(\delta J_{yx}+\delta J_{xy})\frac{\frac{eV}{2T}}{\tanh\frac{eV}{2T}} & -2\left(J_\perp \delta J_{xx}\frac{\frac{eV}{2T}}{\tanh\frac{eV}{2T}} + J_z\delta J_{zz}\right) & J_\perp \delta J_{zy}\frac{\frac{eV}{2T}}{\tanh\frac{eV}{2T}} + J_z\delta J_{yz} \\ J_\perp \delta J_{zx}\frac{\frac{eV}{2T}}{\tanh\frac{eV}{2T}} + J_z\delta J_{xz} & J_\perp \delta J_{zy}\frac{\frac{eV}{2T}}{\tanh\frac{eV}{2T}} + J_z\delta J_{yz} & -2J_\perp(\delta J_{xx}+\delta J_{yy})\frac{\frac{eV}{2T}}{\tanh\frac{eV}{2T}} \end{pmatrix} \mathrm{Re}C_{zz}, \tag{S17}$$

and

$$\mathbf{c}^{(0)} = \mathbf{z}\frac{eV}{2T}J_\perp(T,V)^2 \mathrm{Re}C_{zz}, \quad \mathbf{c}^{(1)} = \frac{eV}{2T}J_\perp(T,V)\begin{pmatrix} -\delta J_{zx}(T,V) \\ -\delta J_{zy}(T,V) \\ \delta J_{xx}(T,V) + \delta J_{yy}(T,V) \end{pmatrix} \mathrm{Re}C_{zz}. \tag{S18}$$

Here we also used the property $\mathrm{Im}C_{xy} = \mathrm{Re}C_{xx}\tanh\frac{eV}{2T}$ [see Eqs. (S5) and (S8)]. Likewise, as we show in Sec. SM4 A, the terms proportional to $\mathrm{Re}C_{xy}$ in $\mathbf{h}$, Eq. (S16), renormalize the exchange couplings $J_{iz}$. Therefore, we can write $\mathbf{h}$ in terms of running couplings as

$$\mathbf{h}^{(0)} = \mathbf{z}\frac{1}{2}eV\rho J_z(T,V), \quad \mathbf{h}^{(1)} = \frac{1}{2}\rho eV\begin{pmatrix} \delta J_{xz}(T,V) \\ \delta J_{yz}(T,V) \\ \delta J_{zz}(T,V) \end{pmatrix}. \tag{S19}$$

To find the steady state spin polarization $\langle \mathbf{S} \rangle$, we set the right-hand side to zero in Eq. (S11). Inserting Eqs. (S17)–(S19) in it allows us to express $\langle \mathbf{S} \rangle$ in terms of the running couplings, in the limit of weak anisotropy. Up to corrections of order $\delta J^2$, we find

$$\langle S_z \rangle = \frac{1}{2}\tanh\frac{eV}{2T}, \tag{S20}$$

$$\langle S_x \rangle = \frac{h_z^{(0)}\left(h_x^{(1)} + \frac{\delta J_{zy}}{J_\perp}\gamma_{zz}^{(0)} - \gamma_{zy}^{(1)}\right) + \tau_K^{-1}\left(h_y^{(1)} - \frac{\delta J_{zx}}{J_\perp}\gamma_{zz}^{(0)} + \gamma_{zx}^{(1)}\right)}{h_z^{(0)2} + \tau_K^{-2}}\frac{1}{2}\tanh\frac{eV}{2T}, \tag{S21}$$

$$\langle S_y \rangle = \frac{h_z^{(0)}\left(h_y^{(1)} - \frac{\delta J_{zx}}{J_\perp}\gamma_{zz}^{(0)} + \gamma_{zx}^{(1)}\right) - \tau_K^{-1}\left(h_x^{(1)} + \frac{\delta J_{zy}}{J_\perp}\gamma_{zz}^{(0)} - \gamma_{zy}^{(1)}\right)}{h_z^{(0)2} + \tau_K^{-2}}\frac{1}{2}\tanh\frac{eV}{2T}. \tag{S22}$$

We denote $\tau_K^{-1} = \gamma_{xx}^{(0)} = (\frac{\frac{eV}{2T}}{\tanh\frac{eV}{2T}}J_\perp^2 + J_z^2)\mathrm{Re}C_{zz}$, which is the notation used in the main text; The variable $x$ in Eq. (14) of the main text is obtained from the ratio $h_z^{(0)}/\tau_K^{-1}$. Note that $\langle S_z \rangle$ is non-zero even without $\delta J$, while the averages $\langle S_{x,y} \rangle$ start from first order in $\delta J$. Also note that $\langle S_{x,y} \rangle$ are non-perturbative in the (bare) isotropic exchange tensor; they lead to physics that were missed by the crude Fermi Golden Rule derivation, Eq. (7), in the introduction of the main text.

## SM3. DERIVATION OF EQ. (10) OF THE MAIN TEXT

Inserting Eq. (S3) [Eq. (9) in the main text] into Eq. (8) of the main text and expressing the result in terms of running couplings (see Sec. SM4 A) leads to

$$\langle \delta I \rangle = \frac{\pi}{2} \rho^2 eT \left( [\delta J_{yz}(T,V)^2 + \delta J_{xz}(T,V)^2] \langle S_z \rangle - J_z(T,V) [\delta J_{yz}(T,V) \langle S_y \rangle + \delta J_{xz}(T,V) \langle S_x \rangle] \right)$$

$$+ \frac{\pi}{4} \rho^2 e^2 V \sum_{i=x,y} \left( \frac{1}{2} \delta J_{zi}(T,V)^2 + \delta J_{zi}(T,V) J_\perp(T,V) \frac{\langle S_i \rangle}{\tanh \frac{eV}{2T}} \right)$$

$$+ \frac{\pi}{4} \rho^2 e^2 V |\delta J_{++}(T,V)|^2 \left( \frac{1}{2} + \frac{\langle S_z \rangle}{\tanh \frac{eV}{2T}} \right) + \frac{1}{2} \rho e^2 V \left( \delta J_{yz}(T,V) \langle S_x \rangle - \delta J_{xz}(T,V) \langle S_y \rangle \right).$$

(S23)

This equation expresses the current in terms of steady-state values of the local-moment spin polarization $\langle \mathbf{S} \rangle$.

Using the steady-state values (S20)-(S22) in Eq. (S23), and expressing them with the help of Eqs. (S17) and (S19) leads to Eqs. (10)–(11) of the main text. For example, the denominator in the function $R$, Eq. (11), is precisely obtained from the denominator in Eqs. (S21), (S22).

## SM4. THE RENORMALIZATION GROUP EQUATIONS FOR THE EXCHANGE COUPLINGS

In this section we will consider the Hamiltonian

$$H = (2\pi)^{-1} v \int dx [\Pi^2 + (\partial_x \varphi)^2] + \sum_{ij} J_{ij}(D) S_i s_j(x_0) \tag{S24}$$

of a helical Luttinger liquid coupled to the magnetic impurity (we use rescaled fields $\varphi$ and $\Pi$ introduced in the main text). Here $D$ is the bare cutoff. We will derive the RG equations for the couplings $J_{ij}$ and solve them in the limits relevant to the main text (*e.g.*, almost isotropic bare couplings).

The RG equations can be derived from poor man scaling [S11–S15], or from an operator product expansion [S16, S17]. In the weak coupling limit $\rho J_{ij} \ll 1$, $1 - K \ll 1$ they read

$$\frac{dJ_{kl}}{d \ln E} = -(1 - \delta_{lz})(1 - K) J_{kl} - \frac{1}{2} \rho \sum_{ijmn} \epsilon_{ijk} \epsilon_{nml} J_{in} J_{jm}. \tag{S25}$$

Let us first consider a fully isotropic initial condition, $\mathbf{J}(D) = \mathbf{1} J_0$. Upon decreasing the running cutoff $E \leq D$, the initially isotropic exchange tensor follows a trajectory $\mathbf{J}_0(E) = \text{diag}(J_\perp, J_\perp, J_z)$ with a uniaxial anisotropy generated by Coulomb interaction, i.e., the first term in Eq. (S25). Along the path, the quantity $[\rho J_z + (1-K)]^2 - (\rho J_\perp)^2$ stays constant. Equivalently, the couplings $J_\perp(E)$ and $J_z(E)$ satisfy Eqs. (5)-(6) of the main text,

$$\frac{dJ_\perp}{d \ln E} = -(1 - K) J_\perp - \rho J_z J_\perp, \tag{S26}$$

$$\frac{dJ_z}{d \ln E} = -\rho J_\perp^2, \tag{S27}$$

which can be derived from Eq. (S25) by using the Ansatz $\mathbf{J}_0(E)$.

Next, we will consider the effect of generic anisotropy. We assume the bare exchange tensor to be weakly anisotropic: $\mathbf{J}(D) = \mathbf{1} J_0 + \delta \mathbf{J}(D)$ with $\delta J_{ij}(D) \ll J_0$. To find how $\delta \mathbf{J}(D)$ flows under RG, we will linearize Eq. (S25) in $\delta \mathbf{J}$. We will find that the anisotropy decreases under RG flow and $\delta \mathbf{J}(E) \ll \mathbf{J}_0(E)$ remains valid. This justifies the linearization. Writing $\mathbf{J} = \mathbf{J}_0 + \delta \mathbf{J}$ in Eq. (S25) and keeping terms to first order in $\delta \mathbf{J}$ gives the following set of decoupled equations

$$\frac{d(\delta J_{xy} + \delta J_{yx})}{d \ln E} = -((1-K) - \rho J_z)(\delta J_{xy} + \delta J_{yx}), \tag{S28}$$

$$\frac{d(\delta J_{xx} - \delta J_{yy})}{d \ln E} = -((1-K) - \rho J_z)(\delta J_{xx} - \delta J_{yy}), \tag{S29}$$

$$\frac{d(\delta J_{zi} + \frac{J_\perp}{J_z} \delta J_{iz})}{d \ln E} = -\left((1-K) - \rho \frac{J_\perp^2}{J_z}\right)(\delta J_{zi} + \frac{J_\perp}{J_z} \delta J_{iz}), \quad i = x, y. \tag{S30}$$

We have given here only the combinations that appear in the backscattering current [see Eq. (10) of the main text]; these are also the irrelevant directions in the case $K = 1$. The directions not shown, for example $\delta J_{xy} - \delta J_{yx}$, are relevant for all values $K \leq 1$. They are relevant since they correspond to either left rotations of the matrix $\mathbf{J}_0$ or

a translation in the RG time (*i.e.*, $\delta \mathbf{J} = \Delta D \cdot \partial_D \mathbf{J}_0$). The latter contribution to the initial $\delta \mathbf{J}$ can be cancelled by shifting the bare cutoff. The former may be cancelled by moving to a basis where the dot spin $\mathbf{S}$ has been rotated by a small angle. For example, the initial condition $(\delta J_{xy} - \delta J_{yx})(D) = 0$ may be imposed by going to a basis where the dot spin has been rotated by an angle $\varepsilon_z = \frac{1}{2}(\delta J_{xy} - \delta J_{yx})/J_0$ about the $z$-axis.

The equations (S26)–(S30) are simple to solve exactly. However, it is more instructive to give the approximate solutions in the limiting cases $E \gg T^*$ and $T_K \ll E \ll T^*$. We shall start from the first, high energy limit.

### A. Solution of Eqs. (S26)–(S30) in the limit $E \gg T^*$

In this limit we can drop the terms second-order-in-$J$ in Eqs. (S26),(S28)–(S30). With the initial condition $J_\perp(D) = J_z(D) = J_0$ this yields

$$J_\perp(E) = (D/E)^{1-K} J_0, \tag{S31}$$

$$J_z(E) = \frac{\rho J_0^2}{2(1-K)}[(D/E)^{2(1-K)} - 1] + J_0, \tag{S32}$$

$$\delta J(E) = (D/E)^{1-K} \delta J(D), \tag{S33}$$

where in the last equation $\delta J$ can be any of $\delta J_{xy} + \delta J_{yx}$, $\delta J_{xx} - \delta J_{yy}$, and $\delta J_{zi} + \frac{J_\perp}{J_z}\delta J_{iz}$. Equation (S33) is behind the power law seen in Eqs. (7) and (14) of the main text. Setting $J_z(T^*) = (1-K)/\rho$ in Eq. (S32) and neglecting the subleading second and third terms of the right-hand side gives immediately Eq. (6) of the main text.

The crude and simple way to account for the bias voltage and temperature dependence of the exchange couplings is to set $E = \max(eV, T)$ in Eqs. (S31)–(S33), *e.g.*,

$$J_\perp(T, V) = (D/\max(eV, T))^{1-K} J_0. \tag{S34}$$

This is however somewhat unsatisfactory since the function $\max(eV, T)$ does not tell what the real behavior near the cross over $T \approx eV$ is. To find the actual cross over function, we use the following approach.

Consider first $K = 1$ with a uniaxially anisotropic exchange tensor, $J(D) = \text{diag}(J_\perp, J_\perp, J_z)$. From Eq. (S13), the Korringa relaxation tensor has then $xx$- and $yy$-components (we write explicitly the bare cutoff $D$)

$$\tau_K^{-1} = \text{Re}C_{zz}[J_\perp^2(D)\frac{\frac{eV}{2T}}{\tanh \frac{eV}{2T}} + J_z^2(D)] + O(J^3), \quad (K = 1). \tag{S35}$$

We also used $\text{Re}C_{xx} = \frac{\tanh \frac{eV}{2T}}{\frac{eV}{2T}}\text{Re}C_{zz}$ (valid for $K = 1$) from Eqs. (S5)–(S6). Here $O(J^3)$ denotes the anticipated corrections that would be found by developing Eq. (S1) to higher order in the exchange: One can imagine developing the perturbation theory to order $J^2$ in Eq. (S3). The second order corrections would lead, for example, to corrections to $\tau_K^{-1}$. We assume that these corrections can be accounted for by using renormalized (running) couplings with a cutoff reduced to $\sim \max(eV, T)$:

$$\tau_K^{-1} = \text{Re}C_{zz}[J_\perp^2(T, V)\frac{\frac{eV}{2T}}{\tanh \frac{eV}{2T}} + J_z^2(T, V)]. \tag{S36}$$

Next, we conjecture that the form (S36) remains valid even when $K \neq 1$, *i.e.*, the Coulomb interaction only renormalizes the exchange couplings in it [and changes $\text{Re}C_{zz}$ by a factor $K$, see Eq. (S6)]. Matching the first term of Eq. (S36) with that of Eq. (S15) leads to the identification

$$J_\perp(T, V) = \sqrt{\frac{\tanh \frac{eV}{2T}}{\frac{eV}{2T}} \frac{\text{Re}C_{xx}}{\text{Re}C_{zz}}} J_0 = \left(\frac{D}{2\pi T}\right)^{1-K} \sqrt{F(\frac{eV}{2T})} J_0. \tag{S37}$$

In the second equality we used Eq. (S5) and introduced the function

$$F(y) = KB(K + i\frac{y}{\pi}, K - i\frac{y}{\pi})\frac{\sinh y}{y}. \tag{S38}$$

Using the large-$y$ asymptote $B(K + i\frac{y}{\pi}, K - i\frac{y}{\pi}) \approx \frac{2\pi}{\Gamma(2K)} e^{-y} \left(\frac{y}{\pi}\right)^{2K-1}$ of the Euler Beta function, one can verify that Eq. (S37) has the same asymptotes as Eq. (S34) for $eV \gg T$ and $eV \ll T$. This supports our conjecture that

Eqs. (S37)-(S38) give the $V$ and $T$ dependence of $J_\perp$ with an accurate cross over function. Finally, one can replace $J_\perp \to \delta J$ in Eq. (S37) to get the bias and temperature dependence of the relevant anisotropic couplings [compare Eqs. (S31) and (S33)]. This leads to Eq. (12) of the main text.

Similarly, comparing Eq. (S16) [the bare exchange field with corrections $\sim J^2 \mathrm{Re} C_{xy}$, Eq. (S7)] to Eq. (S32) leads to the expression (S19) [the fully renormalized exchange field].

### B. Solution of Eqs. (S26)–(S30) in the limit $T_K \ll E \ll T^*$

At temperatures $T \ll T^*$ we have $1 - K \ll \rho J_z, \rho J_\perp$. The Equations (S26)–(S27), can be then approximated by

$$\frac{dJ_\perp}{d \ln E} = -\rho J_z J_\perp \quad \frac{dJ_z}{d \ln E} = -\rho J_\perp^2, \tag{S39}$$

From these equations it follows that $J_z(E) \approx J_\perp(E)$ for $E \ll T^*$. We can then approximate Eqs. (S28)–(S30) by

$$\frac{d\delta J}{d \ln E} = \rho J_z \delta J \tag{S40}$$

where $\delta J$ can be any of $\delta J_{xy} + \delta J_{yx}$, $\delta J_{xx} - \delta J_{yy}$, and $\delta J_{zi} + \delta J_{iz}$. Solving Eq. (S40) with the "initial condition" $\rho J_z(T^*) = 1 - K$ is straightforward and leads to the logarithmic factor in Eq. (17) of the main text.

## SM5.  THE TEMPERATURE DEPENDENCE OF $J_{\mathrm{eff}}(T)$ AND $b(T)$

Let us first discuss the coefficient $J_{\mathrm{eff}}$, defined as [see below Eq. (10) in the main text]

$$J_{\mathrm{eff}}(T) = \frac{J_\perp(T,0)^2 + J_z(T,0)^2}{J_z(T,0)}. \tag{S41}$$

First, for $T \ll T^*$ we have simply $J_{\mathrm{eff}} = 2 J_z$ since then $J_z \approx J_\perp$ as discussed in Sec. SM4 B. At $T \gg T^*$ the behavior of $J_{\mathrm{eff}}(T)$ is more complicated. By using Eqs. (S31)–(S32) in Eq. (S41) we find that $J_{\mathrm{eff}}$ changes from $J_{\mathrm{eff}}(D) = 2 J_0$ to $J_{\mathrm{eff}}(T^*) = 3(1-K)/\rho$; Notably $J_{\mathrm{eff}}$ becomes of order $1 - K$ already at a temperature $\sim \sqrt{T^* D}$. This is because $J_\perp$ grows faster than $J_z$: at temperature $\sim \sqrt{T^* D}$ we have $J_\perp \sim \sqrt{J_0(1-K)/\rho}$ while $J_z \sim J_0$.

From above it follows that the *ratio* $J_z/J_{\mathrm{eff}}$ has a non-monotonic $T$-dependence and acquires a minimum $\sim \rho J_0/(1-K)$ at $\sqrt{T^* D}$. This affects the $T$-dependence of $b(T)$ as we will see in the next section.

### A. Estimating the coefficient $b(T)$

The estimates below Eq. (16) of the main text were obtained in the following way: replacing in Eq. (16) the anisotropic bare couplings by their typical values, $\overline{\delta J_{ij} \delta J_{kl}} = \delta J^2 \delta_{ik} \delta_{jl}$ [here $\overline{(\ldots)}$ denotes averaging over an ensemble of exchange matrices], gives

$$b(T) = \frac{2}{3} + \frac{1}{3} \frac{J_z(T,0)}{J_{\mathrm{eff}}(T)}. \tag{S42}$$

The temperature dependence arises from the ratio $J_z(T,0)/J_{\mathrm{eff}}(T)$. As explained in the beginning of Sec. SM5, the ratio has a non-monotonic $T$-dependence: at $T \sim D$ we have $J_z/J_{\mathrm{eff}} \approx 1/2$ and at $T \sim T^*$ we find $J_z/J_{\mathrm{eff}} \approx 1/3$. The minimum is reached at the intermediate temperature $T \sim \sqrt{T^* D}$ at which $J_z/J_{\mathrm{eff}} \approx \rho J_0/(1-K) \ll 1$. Despite the non-monotonicity of $J_z/J_{\mathrm{eff}}$, it never becomes large and thus has little effect on Eq. (S42). Indeed, we have the bounds $2/3 \approx b(\sqrt{T^* D}) \leq b(T) \leq b(D) \approx 5/6$. These bounds were quoted in the main text below Eq. (16).

---

[S1] H.-P. Breuer and F. Petruccione, *The theory of open quantum systems* (Oxford University Press on Demand, 2002).
[S2] T. Giamarchi, *Quantum Physics in One Dimension*, International Series of Monographs on Physics (Clarendon Press, 2003).

[S3] I. Gradshteyn, I. Ryzhik, and A. Jeffrey, *Table of Integrals, Series and Products 5th edn (New York: Academic)* (1994).
[S4] B. L. Altshuler, I. L. Aleiner, and V. I. Yudson, Phys. Rev. Lett. 111, 086401 (2013).
[S5] O. M. Yevtushenko, A. Wugalter, V. I. Yudson, and B. L. Altshuler, EPL (Europhysics Letters) 112, 57003 (2015).
[S6] T. Giamarchi and H. J. Schulz, Phys. Rev. B 37, 325 (1988).
[S7] I. V. Gornyi, A. D. Mirlin, and D. G. Polyakov, Phys. Rev. Lett. 95, 046404 (2005).
[S8] M. Dyakonov, Solid State Communications 92, 711 (1994).
[S9] F. Bloch, Phys. Rev. 70, 460 (1946).
[S10] J. Korringa, Physica 16, 601 (1950).
[S11] P. Anderson, Journal of Physics C: Solid State Physics 3, 2436 (1970).
[S12] D.-H. Lee and J. Toner, Phys. Rev. Lett. 69, 3378 (1992).
[S13] A. Furusaki and N. Nagaosa, Phys. Rev. Lett. 72, 892 (1994).
[S14] J. Maciejko, Phys. Rev. B 85, 245108 (2012).
[S15] E. Eriksson, Phys. Rev. B 87, 235414 (2013).
[S16] J. Cardy, *Scaling and renormalization in statistical physics*, Vol. 5 (Cambridge university press, 1996).
[S17] D. Sénéchal, in *Theoretical Methods for Strongly Correlated Electrons*, CRM Series in Mathematical Physics, edited by D. Sénéchal, A.-M. Tremblay, and C. Bourbonnais (Springer New York, 2004) pp. 139–186.



\end{document}